\documentclass[10pt,twocolumn,letterpaper]{article}

\usepackage[pagenumbers]{iccv} 

\usepackage{amsmath}
\usepackage{amsfonts}
\usepackage{bm}
\usepackage{mathtools}
\PassOptionsToPackage{hyphens}{url}
\usepackage{xcolor}

\DeclarePairedDelimiterX{\infdivx}[2]{(}{)}{%
  #1\;\delimsize\|\;#2%
}
\newcommand{\kldiv}{D_{\mathrm{KL}}\infdivx}
\DeclarePairedDelimiter{\norm}{\lVert}{\rVert}

\usepackage{algorithm}
\usepackage[noend]{algorithmic}

\usepackage{subcaption}
\usepackage{multirow}
\usepackage{booktabs}

\usepackage{graphicx}
\usepackage{makecell}

\definecolor{iccvblue}{rgb}{0.21,0.49,0.74}
\usepackage[pagebackref,breaklinks,colorlinks,allcolors=iccvblue]{hyperref}
\usepackage[accsupp]{axessibility}  

\def\paperID{8774} 
\def\confName{ICCV}
\def\confYear{2025}

\title{An Inversion-based Measure of Memorization for Diffusion Models}

\author{
Zhe Ma, 
Qingming Li, 
Xuhong Zhang, 
Tianyu Du, 
Ruixiao Lin,\\
Zonghui Wang\thanks{Corresponding authors}, 
Shouling Ji, 
Wenzhi Chen\footnotemark[1]\\
Zhejiang University, Hangzhou, China\\
{\tt\small \{mz.rs,liqm,zhangxuhong,zjradty,linruixiao,zhwang,sji,chenwz\}@zju.edu.cn}
}

\begin{document}
\maketitle

\begin{abstract}
The past few years have witnessed substantial advances in image generation powered by diffusion models. However, it was shown that diffusion models are susceptible to training data memorization, raising significant concerns regarding copyright infringement and privacy invasion. This study delves into a rigorous analysis of memorization in diffusion models. We introduce InvMM\footnote{\url{https://github.com/Maryeon/InvMM}}, an inversion-based measure of memorization, which is based on inverting a sensitive latent noise distribution accounting for the replication of an image. For accurate estimation of the measure, we propose an adaptive algorithm that balances the normality and sensitivity of the noise distribution. Comprehensive experiments across four datasets, conducted on both unconditional and text-guided diffusion models, demonstrate that InvMM provides a reliable and complete quantification of memorization. Notably, InvMM is commensurable between samples, reveals the true extent of memorization from an adversarial standpoint and implies how memorization differs from membership. In practice, it serves as an auditing tool for developers to reliably assess the risk of memorization, thereby contributing to the enhancement of trustworthiness and privacy-preserving capabilities of diffusion models.
\end{abstract}

\section{Introduction}
\label{sec:intro}
Diffusion Models~(DMs)~\cite{sohl2015deep,ho2020denoising} have shown impressive capabilities in the generation of images~\cite{rombach2022high,ramesh2022hierarchical}, videos~\cite{videoworldsimulators2024}, 3D point cloud~\cite{luo2021diffusion}, etc. These techniques lay the foundation for commercial systems or communities such as Stable Diffusion~\cite{rombach2022high}, Midjourney~\cite{Midjourney}, DALL$\cdot$E~\cite{ramesh2022hierarchical,openaiDALLE} and Imagen~\cite{saharia2022photorealistic}, which have attracted millions of active users. The popularity of diffusion models stems from their hierarchical denoising process, which offers high stability for training on billions of data~\cite{schuhmann2022laion} and scalability for multimodal conditional generation~\cite{ramesh2022hierarchical,saharia2022photorealistic}.

However, large-scale datasets used to train the prevalent DMs, e.g., the open-source image-caption dataset LAION~\cite{schuhmann2022laion}, are widely acknowledged to contain content that will raise concerns regarding copyright infringement and privacy invasion. For instance, recent reports revealed that LAION could refer to photographers' works without authorization~\cite{technollamaPhotographerSues} and include private medical photographs~\cite{arstechnicaArtistFinds}. With the uncurated data for training, diffusion models are likely to generate content that infringes the creators' rights or exposes private information~\cite{carlini2023extracting,shan2023glaze}. In this work, we focus on memorization in DMs, one of the most critical issues related to training data misuse.

\begin{figure}[t]
    \centering
    \includegraphics[width=0.9\linewidth]{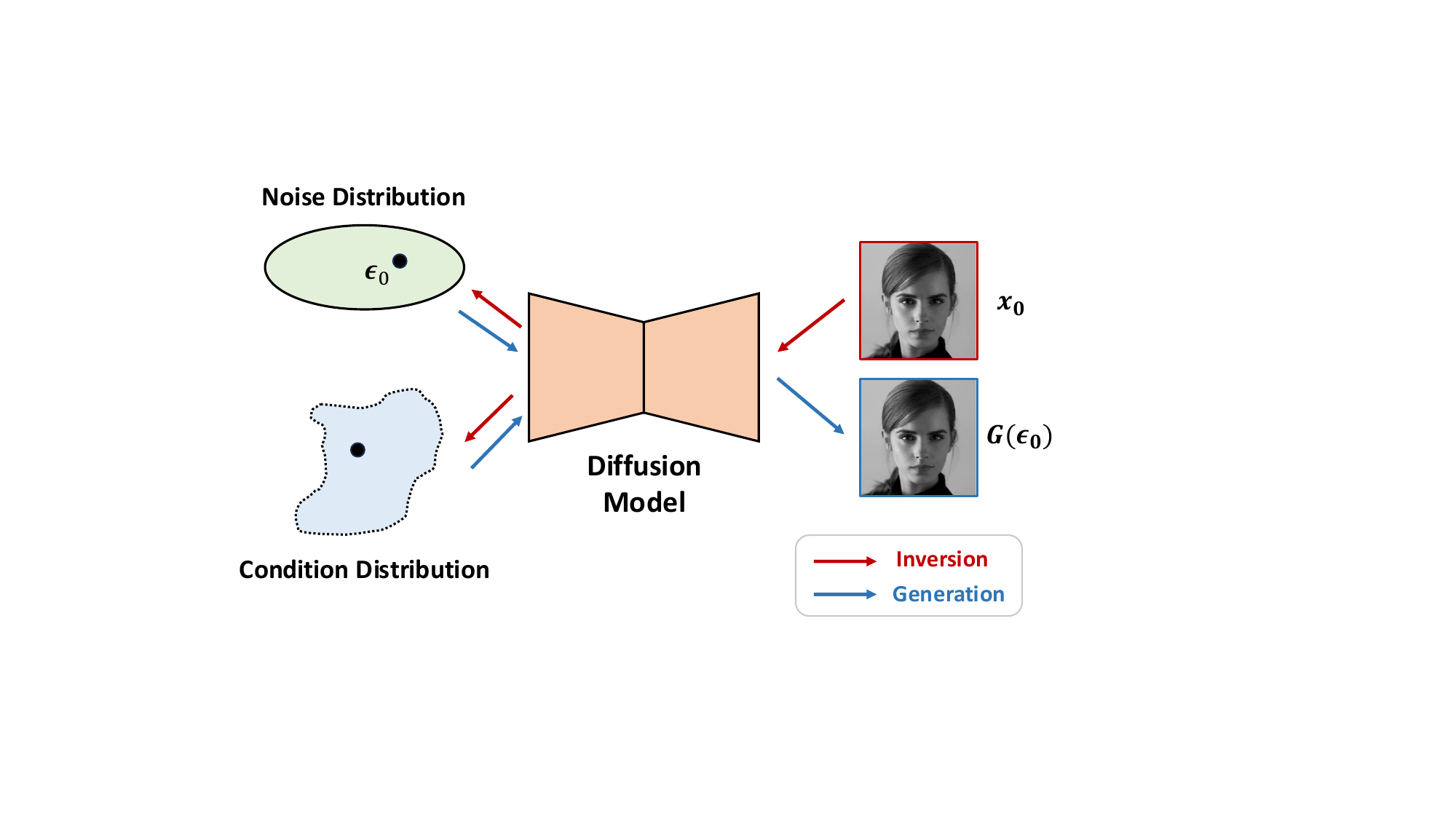}
    \caption{
      Schematic illustration of diffusion model inversion, optionally with conditions.
    }
    \label{fig:inversion}
    \vspace{-2mm}
\end{figure}

Memorization usually manifests itself at sampling time, when a model generates images closely resembling training images.
It can turn into a problem both technically and practically. Technically, heavy memorization indicates failure of generalization to some degree because a probabilistic generative model is supposed to produce novel and diversified images rather than frequently replicate those in the training set.
Practically, memorization might also raise ethical and legal concerns. First, even the model developers are authorized to train their model with protected images, the image owners will never expect their works to be exposed to arbitrary users due to memorization as this would cause indisciplinable dissemination, violating the contextual integrity~\cite{nissenbaum2004privacy}. Second, in past years, text-to-image models have been facing lawsuits for generating derivative images that mimic the style of artists. Compared to derivative generations whose legality is still in pending~\cite{samuelson2023generative}, exact replication of copyrighted images is undisputedly intolerable. Finally, research on privacy protection~\cite{jordon2018pate,packhauser2023generation} has proposed to use synthetic data in place of real data to prevent sharing of private information. For this goal, potential memorization must be carefully circumvented.

Recent studies~\cite{carlini2023extracting,somepalli2023diffusion,somepalli2023understanding,webster} have confirmed the existence of some heavily memorized training data in (text-guided) DMs. They either draw random samples from the model and then detect risky generations of low diversity~\cite{carlini2023extracting}, search for nearest neighbors in the training set~\cite{somepalli2023diffusion,somepalli2023understanding}, or directly identify abnormal prompts~\cite{webster}.
However, conducting a practical analysis of memorization with these approaches still faces challenges. \textit{Transforming the memorization of an image to the presence or absence in samples from the model is not a reliable solution.} While certain images do not appear in a set of samples, it does not demonstrate that they are not highly memorized~\cite{van2021memorization}. For DMs with large latent space, adequate samples should be tested to ensure as complete coverage of the latent space as possible, which could be up to millions of tests~\cite{webster,carlini2023extracting}. Furthermore, things will get much more complicated in conditional DMs, e.g, text-to-image generation. The additional conditions expand the input space of DMs, making the memorization of an image also depend on the specific condition. Existing studies exploit training captions to discover highly memorized training images or perform memorization detection~\cite{wen2024detecting,ren2024unveiling}. However, \textit{attributing the memorization risk of an image to its coupled training caption is incomplete.} It still remains unknown that whether the other images not triggered by their coupled captions are also to a large extent of memorization and to be triggered by potential prompts.


Our contributions are threefold.
(1) To overcome the challenges, we propose an Inversion-based Measure of Memorization~(InvMM) for DMs, a continuous instance-level measure that takes an image as input, outputs a scaler quantifying its level of memorization. 
As shown in \cref{fig:inversion}, the inversion process locates a sensitive latent noise distribution and optionally another condition distribution for the target image, such that generating with noise and condition drawn from the inverted distributions will produce samples exactly the same as the target image. 
InvMM is then defined as the minimum KL divergence of the sensitive noise distribution to the standard Gaussian distribution.
Compared to existing methods, InvMM is more reliable and complete in that it proactively searches for all potential sensitive inputs and does not rely on limited test cases from the training set~(e.g., captions).
(2) To calculate the measure, we design an adaptive optimization algorithm motivated by the TCP congestion control protocol, which is demonstrated to be effective to alleviate overestimate and underestimate of memorization.
(3) Through comprehensive experiments across four datasets, encompassing both unconditional and text-guided DMs, we present the superiority of InvMM in analyzing memorization. Compared to alternative memorization measures, detection metrics and membership inference metrics, InvMM is commensurable between samples, reflects the true memorization level from an adversarial perspective and clarifies the fundamental difference between memorization and membership. As an auditing tool, it has great potential in discovering highly memorized images, validating safety against memorization for any image, determining appropriate training settings, etc.
\section{Related work}
\subsection{Memorization of generative models}
Memorization has raised wide attention on various generative models, including GAN~\cite{goodfellow2014generative}, VAE~\cite{kingma2013auto}, language models~\cite{NEURIPS2020_1457c0d6} and DM~\cite{sohl2015deep}. There have been studies on training algorithms~\cite{salimans2016improved,arjovsky2017wasserstein,ho2020denoising} and evaluation protocols~\cite{gulrajanitowards} to improve the generalization ability of generative models that would potentially circumvent memorization. Empirical and theoretical studies also find that dataset complexity and size~\cite{feng2021gans,somepalli2023diffusion}, training time~\cite{vaishnavh2018theoretical}, training data duplication~\cite{carlini2023quantifying,webster2023duplication} and model capacity~\cite{arora2017gans,carlini2023quantifying} are influential factors of memorization. Several works~\cite{vaishnavh2018theoretical,webster2019detecting} focus on the latent space~(noise space) to analyze memorization. Nagarajan et al.~\cite{vaishnavh2018theoretical} demonstrate that when trained on sufficient random noise vectors, a generator is guaranteed to memorize training data. Webster et al.~\cite{webster2019detecting} measure overfitting of GANs via the reconstruction error at the optimal noise vector. We take a similar perspective on the latent space to analyze memorization while aim to reach an instance-level measure of memorization rather than reveal overall properties of models. In addition, Van der Burg et al.~\cite{van2021memorization} also propose an instance-level measure of memorization as the changed likelihood of one sample when removing it from the training set. However, the measure is not necessarily comparable between datasets, which is important to understand the true level of memorization.

Specifically for diffusion models, recent research~\cite{carlini2023extracting,somepalli2023diffusion,somepalli2023understanding} demonstrates that text-to-image diffusion models suffer from memorization. Some heavily memorized images are discovered from numerous generated samples by membership inference~\cite{carlini2023extracting} or searching for the most similar training images with image matching models~\cite{somepalli2023diffusion}. Webster~\cite{webster} further provides more efficient attacks to extract training images by identifying abnormal prompts. Subsequent works~\cite{wen2024detecting,ren2024unveiling} focus on prompt detection that will trigger training image replication and memorization mitigation~\cite{chen2024towards} at inference time. In response to the disclosed threats, Vyas et al.~\cite{vyas2023provable} propose a provable copyright protection method to prevent replication of sensitive training images.
Our method improves upon existing methods by being commensurable between different samples, basing on images themselves rather than prompts, and working for any samples~(not necessarily training samples). A recent work by Jiang et al.~\cite{jiang2025imagelevel} also considers prompt-free memorization. However, their method extends Wen et al.'s findings~\cite{wen2024detecting} and is restricted to text-guided diffusion models. Our method is established on the property of latent space thus applies to both unconditional and conditional diffusion models.

\subsection{Inversion of diffusion models}
Inversion techniques in diffusion models are widely studied for image editing~\cite{galimage23,mokady2023null,zhang2023inversion}, copyright protection~\cite{wu2024cgi}, generated content detection~\cite{wang2023dire,cazenavette2024fakeinversion}, etc. Through inversion, the object, style and concept contained in the source images can be compressed in latent noises or input token embeddings~(for text-to-image models). Then the inverted latent noises or input token embeddings are utilized to generate similar images that preserve the desired content.

We leverage inversion techniques to analyze memorization in diffusion models. In contrast, we invert a sample to a distribution consisting of all feasible latent vectors rather than to a single vector~\cite{galimage23,mokady2023null,wu2024cgi,wang2023dire,cazenavette2024fakeinversion}. We also adopt more efficient training objective for inversion, without the need to iteratively invert a specific sampling procedure~\cite{mokady2023null,wu2024cgi,wang2023dire,cazenavette2024fakeinversion}.
\section{Background}
\label{sec:bg_diffusion_model}

    Basically, DMs~\cite{sohl2015deep,ho2020denoising} are likelihood-based generative models that learn to maximize the likelihood $p_{\bm{\theta}}(\bm{x}_0)$ of observed data $\bm{x}_0$. Directly calculating $p_{\bm{\theta}}(\bm{x}_0)$ is intractable for continuous data such as image. By contrasting to a sequence of latent variables $\bm{x}_{1:T}$, the log-likelihood $\log p_{\bm{\theta}}(\bm{x}_0)$ can be maximized indirectly by maximizing its variational lower bound~\cite{KingmaW13,luo2022understanding}:
\begin{align}
    \log p_{\bm{\theta}}(\bm{x}_0) &= \log \int p_{\bm{\theta}}(\bm{x}_{0:T})d\bm{x}_{1:T}\\
    &\geq \mathbb{E}_{q(\bm{x}_{1:T}|\bm{x}_0)} \left[  \log \frac{p_{\bm{\theta}}(\bm{x}_{0:T})}{q(\bm{x}_{1:T}|\bm{x}_0)} \right] \label{eq:var_low_bound}
\end{align}
Assume both transitions $p_{\bm{\theta}}(\bm{x}_{0:T})$ and $q(\bm{x}_{1:T}|\bm{x}_0)$ are Markov chains, i.e., $q(\bm{x}_{1:T}|\bm{x}_0)=\prod_{t=1}^T q(\bm{x}_t|\bm{x}_{t-1})$, $p_{\bm{\theta}}(\bm{x}_{0:T})=p(\bm{x}_T)\prod_{t=1}^T p_{\bm{\theta}}(\bm{x}_{t-1}|\bm{x}_t)$, the variational lower bound in \cref{eq:var_low_bound} expands as:
\begin{align}
    &\log p_{\bm{\theta}}(\bm{x}_0)\geq\mathbb{E}_{q(\bm{x}_1|\bm{x}_0)} \left[ \log p_{\bm{\theta}}(\bm{x}_0|\bm{x}_1) \right]\notag\\
    &-\sum_{t=2}^{T} \mathbb{E}_{q(\bm{x}_t|\bm{x}_0)} \left[ \kldiv{q(\bm{x}_{t-1}|\bm{x}_t,\bm{x}_0)}{p_{\bm{\theta}}(\bm{x}_{t-1}|\bm{x}_t)} \right]\label{eq:match_term}\\
    &-\kldiv{q(\bm{x}_T|\bm{x}_0)}{p(\bm{x}_T)} \label{eq:kl_div}
\end{align}
where $p(\bm{x}_T)$ is generally assumed to be a simple distribution such as standard Gaussian $\mathcal{N}(\mathbf{0},\mathbf{I})$.

The transition $q(\bm{x}_{1:T}|\bm{x}_0)$ is not learned but replaced by a manually crafted non-parametric diffusion process. The transition $p_{\bm{\theta}}(\bm{x}_{0:T})$ is the corresponding reverse~(or generation, sampling) process. The diffusion process gradually adds Gaussian noises to the input $\bm{x}_0$ according to a weight schedule $\alpha_{1:T}$:
\begin{gather}
    \bm{x}_t=\sqrt{\alpha_t}\bm{x}_{t-1}+\sqrt{1-\alpha_t}\bm{\epsilon}_{t-1},\bm{\epsilon}_{t-1}\sim\mathcal{N}(\mathbf{0},\mathbf{I})\\
    \bm{x}_t=\sqrt{\bar{\alpha}_t}\bm{x}_0+\sqrt{1-\bar{\alpha}_t}\bm{\epsilon}_0,\bm{\epsilon}_0\sim\mathcal{N}(\mathbf{0},\mathbf{I})
    \label{eq:add_noise}
\end{gather}
where $\bar{\alpha}_t=\prod_{i=1}^t \alpha_i$ decreases over time and finally to almost zero at the last step $T$, such that $q(\bm{x}_T|\bm{x}_0)$ approaches to a standard Gaussian and thus \cref{eq:kl_div} approximately equals to 0.

If the diffusion process is divided into sufficient steps, each reverse transition $p_{\bm{\theta}}(\bm{x}_{t-1}|\bm{x}_t)$ can also be approximated by a Gaussian transformation with learnable parameters that is trained to match the corresponding posterior diffusion transition $q(\bm{x}_{t-1}|\bm{x}_t,\bm{x}_0)$. Omitting further details~\cite{luo2022understanding}, the following simplified objective~\cite{ho2020denoising} as an estimate of the variational lower bound is commonly used for training:
\begin{equation}
    l_{de}= \mathbb{E}_{\bm{\epsilon}_0\sim\mathcal{N}(\mathbf{0},\mathbf{I}),t\sim\mathcal{U}\{1,T\}} \left [ \norm{\bm{\epsilon}_0-\bm{\epsilon}_{\bm{\theta}}(\bm{x}_t,t)}_2^2 \right ]
    \label{eq:epsilon_loss}
\end{equation}
where $\bm{\epsilon}_{\bm{\theta}}$ is a neural network that takes the noisy data $\bm{x}_t$ and timestep $t$ as input and predicts the added noise $\bm{\epsilon}_0$. After training, the vanilla sampling procedure starts with a random Gaussian noise $\bm{x}_T\sim\mathcal{N}(\mathbf{0},\mathbf{I})$ and removes the predicted noise stepwise.
However, it is extremely slow to generate an image as it must invoke the network $\bm{\epsilon}_{\bm{\theta}}$ for $T$ times (typically 1000 steps). To mitigate the problem, a variety of efficient sampling algorithms are proposed, such as DDIM sampler~\cite{songdenoising} and PLMS sampler~\cite{liupseudo}.

\section{Inversion-based measure}
\label{sec:method}

Given an image $\bm{x}_0$, we aim to measure to what extent $\bm{x}_0$ is memorized by a diffusion model~(DM). We propose to invert the image into a latent noise distribution, such that noise vectors drawn from it will lead to replication of the image $\bm{x}_0$.
Our memorization measure is established in this inversion process.
In this section, we introduce a sensitivity definition of the inverted noise distribution~(\cref{sec:sensitivity}), elaborate the memorization measure~(\cref{sec:measure}) and give an adaptive algorithm for its calculation~(\cref{sec:adaptive_alg}). Ultimately, we present an example to measure memorization in conditional DM, specifically text-to-image DM~(\cref{sec:text_guided_dm_measure}).

\subsection{Sensitivity of latent distribution}
\label{sec:sensitivity}

The diversity of DM generated content lies in the randomness of latent noise vectors. Starting from different noise vectors, the generated contents generally vary. In the context of memorization, this motivates us to search for \textit{sensitive noise vectors} that will replicate the target image. We then figure out ``how many'' sensitive noise vectors there are and adopt it as a meaningful measure.

Considering the fact that the common latent noise is sampled from a standard Gaussian distribution, we correspondingly assume the sensitive noise vectors follow a distribution $q_{\bm{\varphi}}(\bm{\epsilon}_0|\bm{x}_0)$ with learnable parameters $\bm{\varphi}$.
The sensitivity function $\mathcal{S}(q_{\bm{\varphi}})$ of a noise distribution can be defined as the probability that generated image $G(\bm{\epsilon}_0)$ obtained from random noise $\bm{\epsilon}_0\sim q_{\bm{\varphi}}(\bm{\epsilon}_0|\bm{x}_0)$ replicates the target image:
\begin{equation}
    \mathcal{S}(q_{\bm{\varphi}})=\int \mathbf{1}_{d(G(\bm{\epsilon}_0),\bm{x}_0)\leq \beta} q_{\bm{\varphi}}(\bm{\epsilon}_0|\bm{x}_0) d\bm{\epsilon}_0
\end{equation}
where $\mathbf{1}_v$ is an indicator function that equals 1 if v is true and 0 otherwise, $d(\cdot,\cdot)$ is a distance metric between two images and $\beta$ is an appropriate threshold.

The choice of the distance metric depends on specific applications. In applications that accept derivative generation, a strict metric that only detects eidetic generation can be utilized. While under some circumstances even similar generation should be circumvented, a looser metric is needed. In this work, we consider exact replication that the generated image and target image are exactly the same. This consideration to some extent reduces the ambiguity of determining the similarity of two images. Following the investigation of Somepalli et al.~\cite{somepalli2023diffusion}, we implement the metric as the Euclidean distance between embeddings extracted by the pre-trained copy detection model SSCD~\cite{pizzi2022self}. 

\subsection{Quantify memorization}
\label{sec:measure}

Based on the definition of sensitivity, we propose an \textbf{I}nversion-based \textbf{M}easure of \textbf{M}emorization~(InvMM), as the best achievable average normality of $q_{\bm{\varphi}}(\bm{\epsilon}|\bm{x})$ under the condition that $\mathcal{S}(q_{\bm{\varphi}})=1$:
\begin{align}
    &\min_{\bm{\varphi}} \frac{1}{N} \kldiv{q_{\bm{\varphi}}(\bm{\epsilon}_0|\bm{x})}{p(\bm{\epsilon}_0)} \notag\\
    &\mathrm{s.t.}~\mathcal{S}(q_{\bm{\varphi}})=1
    \label{eq:measure}
\end{align}
where $N$ is the dimensionality of latent noise space and $p(\bm{\epsilon}_0)$ is the prior distribution used in DM, which is often selected to be standard Gaussian. The normality of inverted noise distribution is the KL divergence to the prior. A lower value of InvMM means greater degree of memorization.

\begin{figure}[tbp]
    \centering
    \begin{subfigure}[t]{0.25\linewidth}
        \centering
        \includegraphics[width=\linewidth]{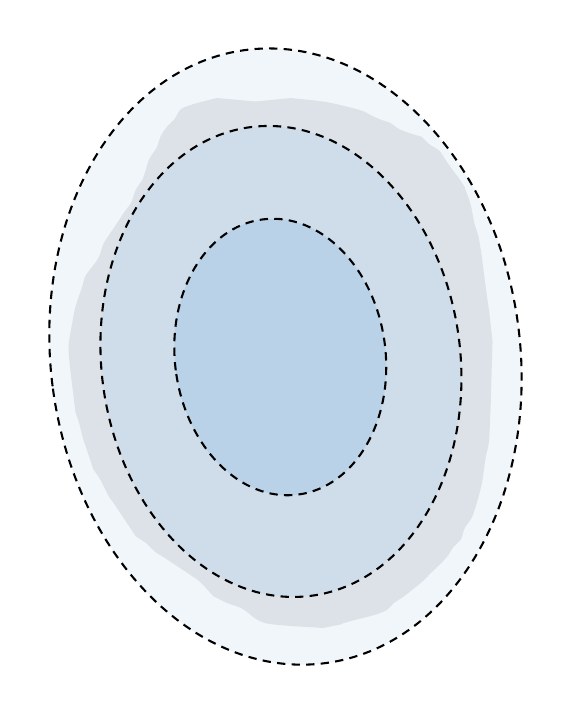}
        \caption{Optimal}
    \end{subfigure}
    \begin{subfigure}[t]{0.25\linewidth}
        \centering
        \includegraphics[width=\linewidth]{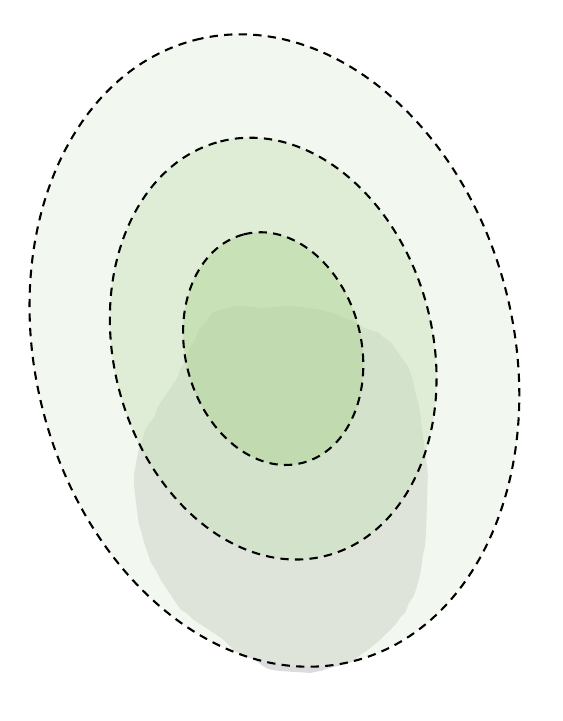}
        \caption{Overestimate}
        \label{fig:overestimate}
    \end{subfigure}
    \begin{subfigure}[t]{0.25\linewidth}
        \centering
        \includegraphics[width=\linewidth]{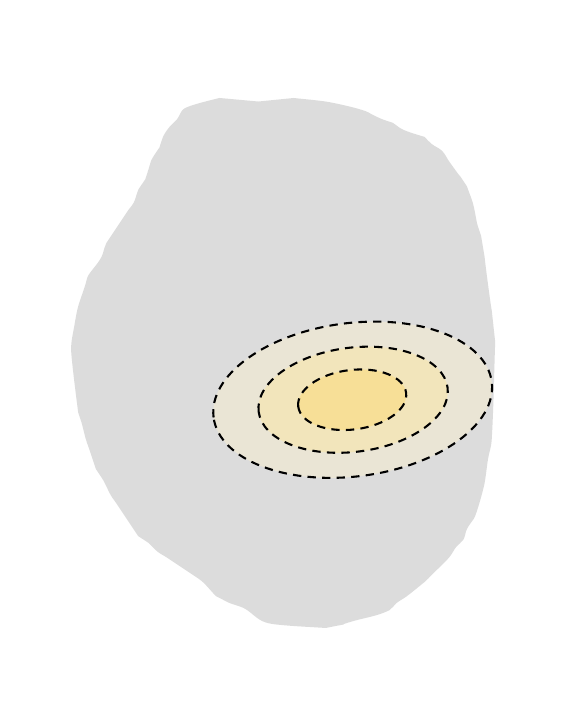}
        \caption{Underestimate}
        \label{fig:underestimate}
    \end{subfigure}
    \caption{Illustration of our memorization measure. The gray blocks indicate the distribution of sensitive latent noise vectors.}
    \label{fig:optimal_inversion}
\end{figure}

\textbf{Explaining InvMM}. Partitioning the normality gives us a more intuitional understanding of the measure. Notice that
\begin{gather}
    \kldiv{q_{\bm{\varphi}}(\bm{\epsilon}_0|\bm{x}_0)}{p(\bm{\epsilon}_0)}= \notag\\
    \mathrm{H}(q_{\bm{\varphi}}(\bm{\epsilon}_0|\bm{x}_0),p(\bm{\epsilon}_0))-\mathrm{H}(q_{\bm{\varphi}}(\bm{\epsilon}_0|\bm{x}_0))
\end{gather}
where $\mathrm{H}(q_{\bm{\varphi}}(\bm{\epsilon}_0|\bm{x}_0),p(\bm{\epsilon}_0))$ is the cross entropy between the inverted distribution and the prior distribution, $\mathrm{H}(q_{\bm{\varphi}}(\bm{\epsilon}_0|\bm{x}_0))$ is the differential entropy of the inverted distribution. Minimizing the normality thus encourages a distribution as close to the prior distribution as possible, meanwhile with the largest entropy.
For the cross entropy, following the Jensen's inequality, $-\mathrm{H}(q_{\bm{\varphi}}(\bm{\epsilon}_0|\bm{x}_0),p(\bm{\epsilon}_0))=\mathbb{E}_{q_{\bm{\varphi}}(\bm{\epsilon}_0|\bm{x}_0)}[\log p(\bm{\epsilon}_0)]\leq \log \int q_{\bm{\varphi}}(\bm{\epsilon}_0|\bm{x}_0)p(\bm{\epsilon}_0)d\bm{\epsilon}_0$. The term $q_{\bm{\varphi}}(\bm{\epsilon}_0|\bm{x}_0)p(\bm{\epsilon}_0)$ is proportional to the probability of a random noise $\bm{\epsilon}_0$, with probability density $q_{\bm{\varphi}}(\bm{\epsilon}_0|\bm{x}_0)$, to be sampled from the prior distribution with probability density $p(\bm{\epsilon}_0)$. In the sense, minimizing the cross-entropy improves the probability that sensitive noise vectors regarding an image $\bm{x}_0$ are sampled in the normal usage of DM.
For the differential entropy, maximizing it will produce a noise distribution involving ``as many'' sensitive noise vectors as possible. In the extreme case, suppose a uniform distribution over the entire latent space having the largest entropy, it means that any noise in the space can cause replication of the target training sample.

InvMM requires to achieve the best trade-off~between improving normality and meeting the sensitivity condition. Different images to varied extent of memorization may reach different equilibrium states. \Cref{fig:optimal_inversion} illustrates the optimal solution compared to suboptimal choices.
Overestimate of memorization produces a distribution with good normality but low sensitivity~(\cref{fig:overestimate}). Conservative inversion of the latent distribution perfectly meets the sensitivity condition while underestimates memorization~(\cref{fig:underestimate}).


\subsection{Adaptive algorithm}
\label{sec:adaptive_alg}

We then introduce an adaptive algorithm to solve the inversion problem.

\textbf{Transforming the inversion problem.}
It is intractable to directly calculate the sensitivity condition. For diffusion models, it can be achieved by minimizing the denoising error $l_{de}$~(see \cref{eq:epsilon_loss}) as a proxy. As can be expected, when $l_{de}$ gradually decreases, $\mathcal{S}(q_{\bm{\varphi}})$ will approach 1.

Moreover, denote the average KL divergence in \cref{eq:measure} by $l_{kl}$, the optimization problem can be solved by Lagrangian method:
\begin{equation}
    \min_{\bm{\varphi},\lambda} l_{de}(\bm{x}_0;\bm{\varphi})+\lambda l_{kl}(\bm{x}_0;\bm{\varphi}), \lambda\geq 0
    \label{eq:reweighted_elbo}
\end{equation}
where $\lambda\geq 0$ is a multiplier. We attach the multiplier to $l_{kl}$ to ensure adequate optimization of the denoising error $l_{de}$.

Different from the standard $\bm{\epsilon}_0$-prediction denoising error, we adopt $\bm{x}_0$-prediction denoising error, which shows more stable inversion efficacy in practice:
\begin{align}
    l_{de}&= \mathbb{E}_{q_{\bm{\varphi}}(\bm{\epsilon}_0|\bm{x}_0),t} \left [ \norm{\bm{x}_0-\bm{x}_{\bm{\theta}}(\bm{x}_t,t)}_2^2 \right ] \notag\\
    &= \frac{1-\bar{\alpha}_t}{\bar{\alpha}_t}\mathbb{E}_{q_{\bm{\varphi}}(\bm{\epsilon}_0|\bm{x}_0),t} \left [ \norm{\bm{\epsilon}_0-\bm{\epsilon}_{\bm{\theta}}(\bm{x}_t,t)}_2^2 \right ]
    \label{eq:x0_prediction}
\end{align}
where $\bm{x}_t$ is obtained as in~\cref{eq:add_noise} with added noise drawn from $q_{\bm{\varphi}}$ rather than standard Gaussian. 

For the noise distribution $q_{\bm{\varphi}}(\bm{\epsilon}_0|\bm{x}_0)$, we assume it to be a multivariate Gaussian $\mathcal{N}(\bm{\mu},\bm{\sigma}^2)$ with learnable mean $\bm{\mu}$ and diagonal variance $\bm{\sigma}^2$, i.e., $\bm{\varphi}=(\bm{\mu},\bm{\sigma}^2)$. Then InvMM has a closed-form expression $\frac{1}{2N} \sum_i (\bm{\mu}^2_i+\bm{\sigma}^2_i-\log\bm{\sigma}^2_i-1)$. Although the true distribution of sensitive latent noise might not be fully captured by a Gaussian, we choose this modeling for simplicity and find it effective in experiments.

\begin{algorithm}[thbp]
\caption{InvMM: adaptive weight tuning algorithm.}
\label{alg:measure}
\begin{algorithmic}[1]
\REQUIRE Noise prediction network $\bm{\epsilon}_{\bm{\theta}}$, target image $\bm{x}_0$, optimization iteration $T$, observation cycle $C$, weight increment $\delta$, threshold $\xi$, learning rate $\gamma$
\STATE Initialize $\bm{\varphi}\gets (\bm{\mu}\gets \mathbf{0},\bm{\sigma}\gets \mathbf{I}),\lambda\gets 1,l_{de}^p\gets +\infty,\mathrm{success}\gets\mathrm{FALSE}$
\FOR{$i\gets 1$ to $T$}
\STATE $\mathcal{L}\gets l_{de}(\bm{x}_0;\bm{\varphi})+\lambda l_{kl}(\bm{x}_0;\bm{\varphi})$
\STATE $\bm{\varphi}\gets \bm{\varphi}-\gamma\nabla_{\bm{\varphi}}\mathcal{L}$
\IF{$i>1~\textbf{and}~i~\%~C=0$}
\IF{$l_{de}-l_{de}^p<\xi$}
\STATE $\lambda\gets \lambda~/~2$
\ELSE
\STATE $\lambda\gets \lambda+\delta$
\ENDIF
\STATE $l_{de}^p\gets l_{de}$
\IF{$\mathcal{S}(q_{\bm{\varphi}})=1$}
\STATE success$\gets$TRUE
\STATE \textbf{break}
\ENDIF
\ELSE
\STATE $\lambda\gets \lambda+\delta$
\ENDIF
\ENDFOR
\IF{success}
\RETURN $\frac{1}{2N} \sum_i (\bm{\mu}^2_i+\bm{\sigma}^2_i-\log\bm{\sigma}^2_i-1)$
\ELSE
\RETURN $+\infty$
\ENDIF
\end{algorithmic}
\end{algorithm}

\textbf{Adaptive weight tuning.}
To achieve the best trade-off, the weight $\lambda$ should be adjusted appropriately for different samples. However, it cannot be optimized via gradient decent together with $\bm{\varphi}$. As the gradient w.r.t. $\lambda$ is $l_{kl}$, which is always nonnegative, $\lambda$ will gradually decrease to 0 and $l_{kl}$ will be totally ignored. To solve the problem, we propose an algorithm that allows adaptive tuning of $\lambda$ during inversion, motivated by the additive increase/multiplicative decrease scheme in congestion control of TCP protocol, as in Algorithm~\ref{alg:measure}, line 6-9.

Specifically, we monitor the denoising error $l_{de}$ every $C$ steps during the optimization process and divide the weight $\lambda$ by half if no significant improvement of denoising error is observed in the cycle. Otherwise, $\lambda$ slowly increases by a small increment $\delta$ each step. In this way, $\lambda$ is adaptively adjusted for different samples to balance the normality and denoising error. When the normality is too strong to further reduce the denoising error, thus to meet the sensitivity condition, it is immediately down-weighted. Otherwise, we gradually strengthen it for better normality. We also perform early stop~(line 11-13) when the sensitivity condition is reached to avoid unnecessary normality degradation. The sensitivity condition is tested via the Monte Carlo method. Random latent vectors are drawn to generate samples. If the distances between all of them and the target image are lower than the threshold $\beta$, then the inversion is successful. There are cases that only partial samples replicate that target image in the end. We still consider such cases as insignificant memorization and flag them as unsuccessful.

\subsection{Text-to-image DM}
\label{sec:text_guided_dm_measure}

For text-to-image DM, we model the prompt part as a distribution as well and optimize it simultaneously with the noise distribution.
Suppose an input prompt is tokenized into maximum $M$ tokens $\bm{\omega}=[ \bm{\omega}_1, \bm{\omega}_2, ..., \bm{\omega}_M ]$. Motivated by Guo et al.~\cite{guo2021gradient}, we learn a $|\mathcal{V}|$-dimensional categorical distribution over the vocabulary $\mathcal{V}$ for each token $\bm{\omega}_i$. Let token $\bm{\omega}_i$ follows a categorical distribution with probabilities $\bm{\pi}_{i,j}$, where $\sum_{j=1}^{|\mathcal{V}|} \bm{\pi}_{i,j}=1$. $\bm{\pi}_{i,j}$ can be parameterized and optimized.

Because sampling from categorical distribution is non-differentiable, we adopt the Gumbel-Softmax reparameterization~\cite{jang2016categorical} to obtain smoothed sample $ \tilde{\bm{\omega}}_i$. Given the smoothed token $\tilde{\bm{\omega}}_i$, it can be converted to token embedding as a linear combination of vocabulary embeddings $e(\tilde{\bm{\omega}}_i)=\sum_{j=1}^{|\mathcal{V}|} \tilde{\bm{\omega}}_{i,j}e(j)$, where $e(j)$ is the pretrained token embedding of the $j$-th token. A proper choice of the prompt distribution is critical for inversion on text-to-image DMs. We provide a detailed analysis in Appendix \cref{sec:text_to_image_inversion}.
\section{Experiment}
In this section, we conduct a series of experiments from different perspectives to demonstrate the validity of InvMM. First, we conduct controlled studies to verify the effectiveness of the adaptive algorithm~(\cref{sec:exp_adaptive_alg}). Based on this, we show that InvMM can reasonably elucidate the influence of various factors on memorization~(\cref{sec:factor}) and identify highly memorized images~(\cref{sec:detect}). Finally, we discuss how memorization differs from membership~(\cref{sec:membership}).

\subsection{Experiment setup}
\textbf{Dataset.}
We use four datasets for evaluation, i.e., CIFAR-10~\cite{krizhevsky2009learning}, CelebAHQ~\cite{karras2018progressive}, FFHQ~\cite{karras2019style} and LAION-Aesthetics V2~\cite{schuhmann2022laion}. For subsequent experiments, we construct four randomly sampled subsets of CelebAHQ and FFHQ: CelebAHQ-250, CelebAHQ-2.5k, FFHQ-600 and FFHQ-6k. We ensure that the larger subset includes the smaller subset. The LAION dataset is extremely large. To keep computational costs manageable, we collect a subset of images. We first include 76 heavily memorized images discovered by Webster~\cite{webster}. We call this set of images the \textit{confirmed} set. In addition, we collected another 50 images with high duplication rates according to the statistics from Webster et al~\cite{webster2023duplication}. This set of suspects, named as the \textit{suspicious} set, are assured to be disjoint with the \textit{confirmed} set. Images in the \textit{confirmed} set can be triggered by their training captions while those in the \textit{suspicous} set cannot. Finally, another 50 randomly sampled images are used for comparison, referred to as the \textit{normal} set.

\textbf{Diffusion Models.}
Following the implementations of DDPM~\cite{ho2020denoising} and Latent Diffusion Model~(LDM)~\cite{rombach2022high}, we train DDPM on CIFAR-10 and LDM on the subsets of CelebAHQ and FFHQ. Pre-trained LDMs on the whole CelebAHQ and FFHQ are also used. For text-to-image generation, we utilize Stable Diffusion~(SD) v1.4, v2.1 and v3.5 trained on LAION.

\textbf{Inversion.}
The experiment setting and additional results are deferred to Appendix~\cref{sec:additional_exp}.

\subsection{Adaptive algorithm}
\label{sec:exp_adaptive_alg}

\begin{figure}[t]
    \centering
    \includegraphics[width=0.9\linewidth]{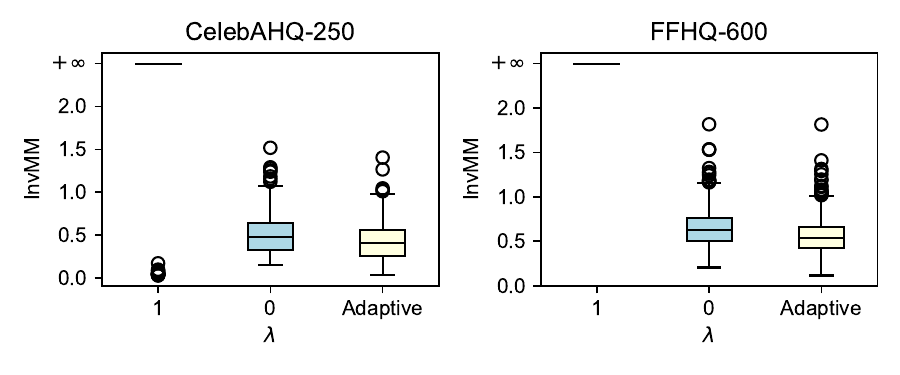}
    \caption{
    The influence of weight $\lambda$ on measuring memorization.
    }
    \label{fig:lambda}
    \vspace{-3mm}
\end{figure}

\begin{figure}[tbp]
    \centering
    \begin{subfigure}[t]{0.4\linewidth}
        \centering
        \includegraphics[width=\linewidth]{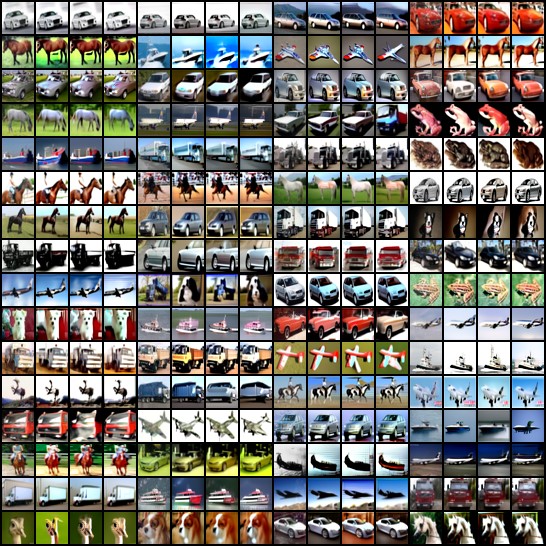}
        \caption{CIFAR-10~(DDPM)}
    \end{subfigure}
    \begin{subfigure}[t]{0.4\linewidth}
        \centering
        \includegraphics[width=\linewidth]{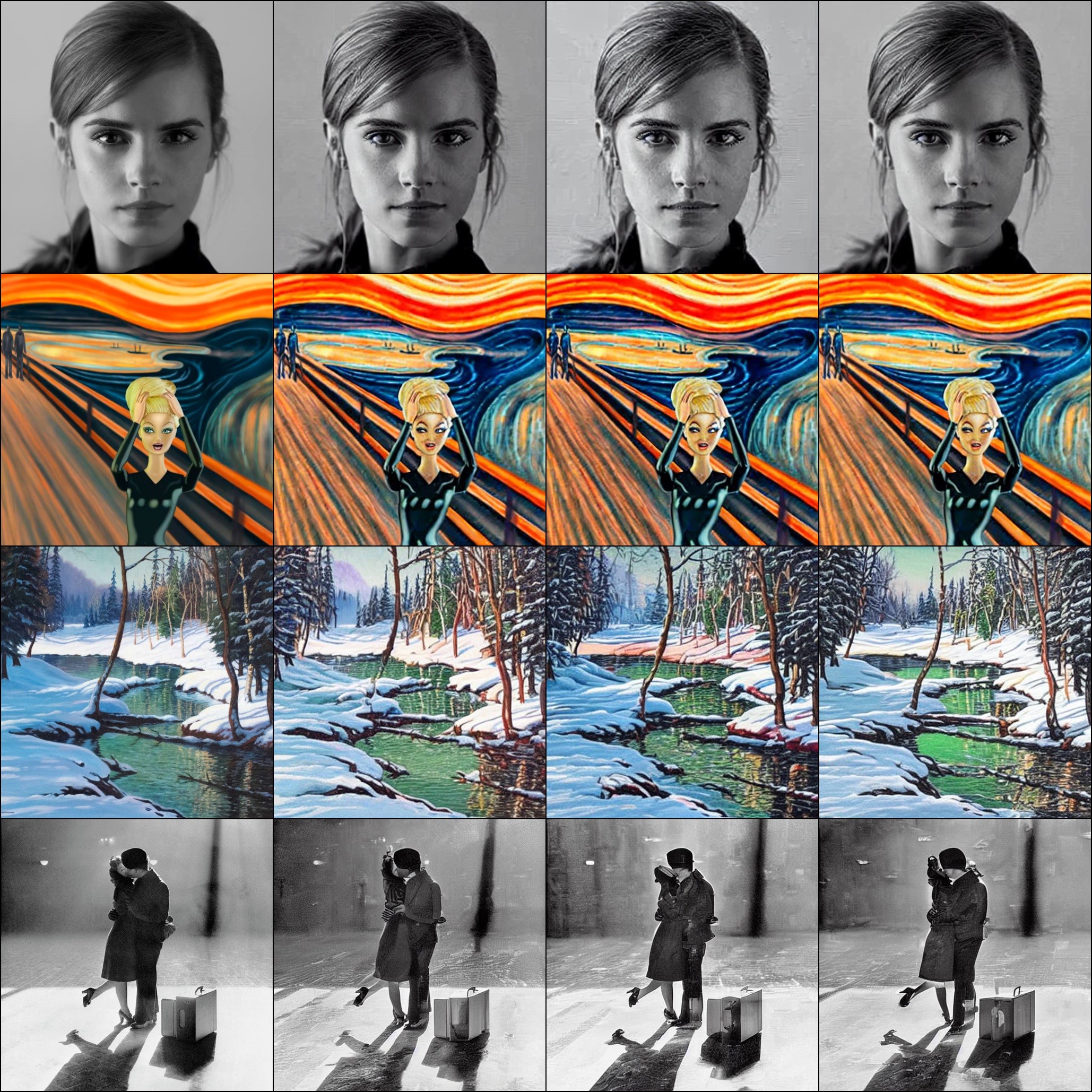}
        \caption{LAION~(SD)}
    \end{subfigure}
    \begin{subfigure}[t]{0.4\linewidth}
        \centering
        \includegraphics[width=\linewidth]{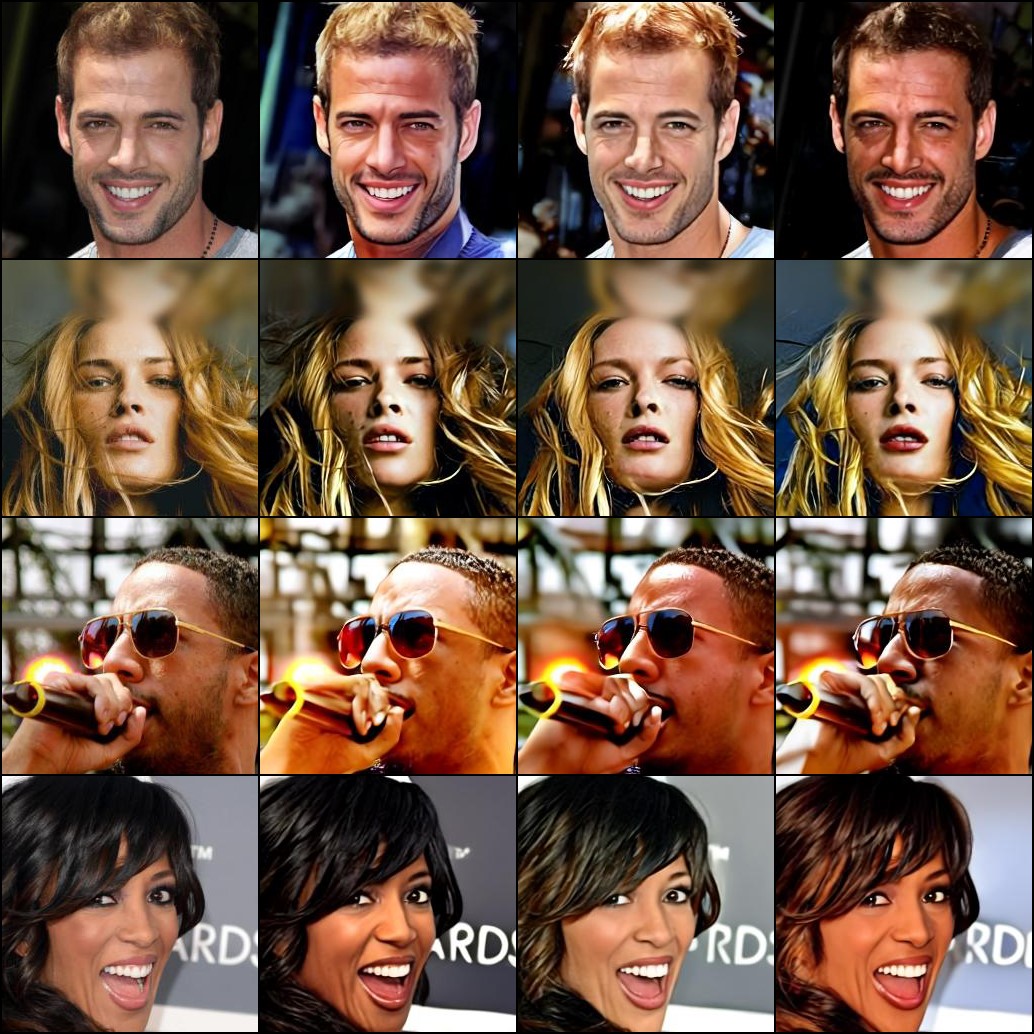}
        \caption{CelebAHQ~(LDM)}
    \end{subfigure}
    \begin{subfigure}[t]{0.4\linewidth}
        \centering
        \includegraphics[width=\linewidth]{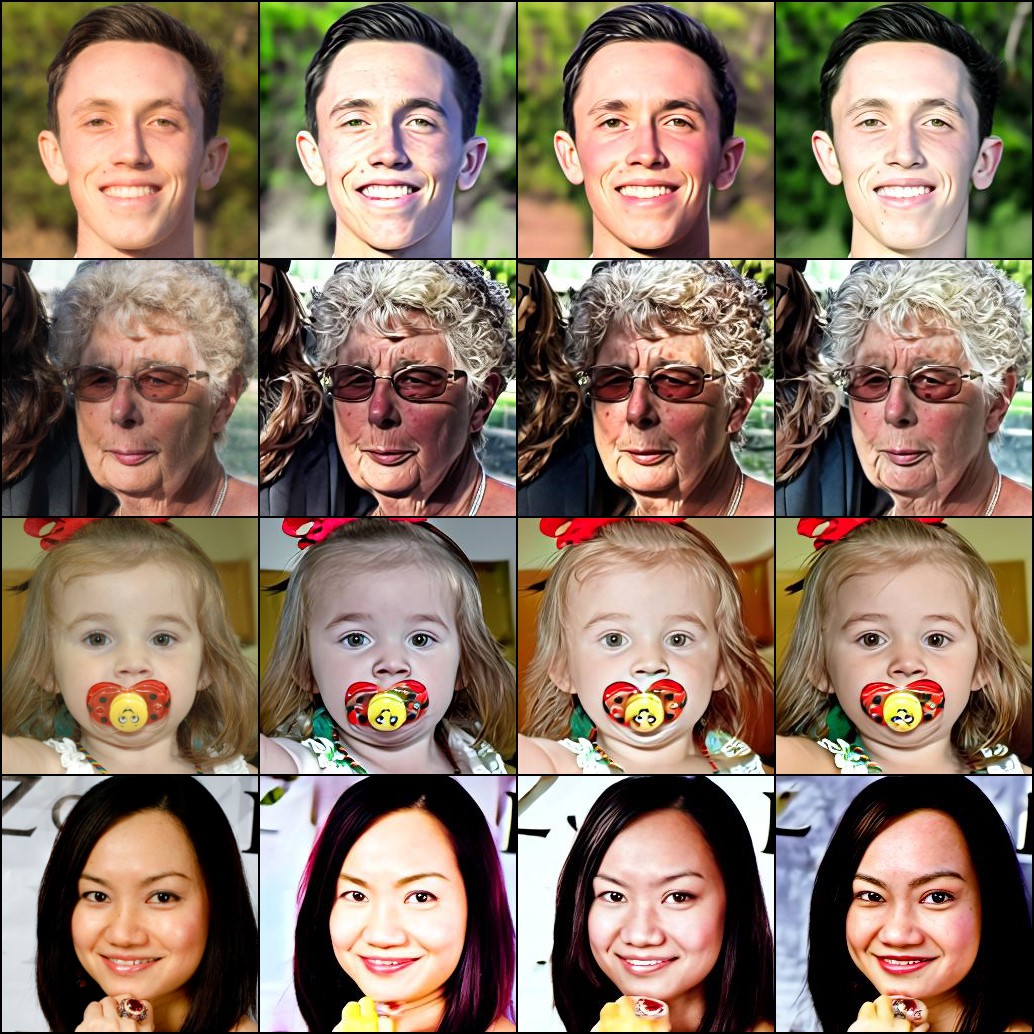}
        \caption{FFHQ~(LDM)}
    \end{subfigure}
    \caption{Random samples generated using latent noise vectors drawn from inverted distributions. The original training images are in the first column within each block.}
    \label{fig:inversion_examples}
    \vspace{-3mm}
\end{figure}

The core design of \cref{alg:measure} is the dynamic reweighting of the two loss items. To verify its effectiveness, we compare to fixed weights $\lambda=1$ and $\lambda=0$. $\lambda=1$ equals DM's original training objective~(\cref{eq:match_term,eq:kl_div}) and $\lambda=0$ totally ignores the normality constraint.

A comparison of the resulting memorization score under different settings can be observed in \cref{fig:lambda}.
Fixing $\lambda$ to 1 cannot successfully invert most images: 92.4\% and 99.7\% images of CelebAHQ and FFHQ result in infinite memorization scores. In contrast, both $\lambda=0$ and adaptive reweighting successfully invert all images. However, Fixing $\lambda$ to 0 leads to larger resulting memorization scores, which indicates underestimation of the extent of memorization.

Figure~\ref{fig:inversion_examples} showcases the generation results with the learned distributions. With the adaptive algorithm, we successfully perform inversion across various datasets and DMs that the randomly generated samples closely resemble the training images. The results form the premise of InvMM that memorization of samples is measured in the state they are made replicable.

\subsection{Influence factors}
\label{sec:factor}

Both theoretical analysis and empirical investigation point out that dataset complexity~\cite{feng2021gans,somepalli2023understanding}, dataset size~\cite{feng2021gans,somepalli2023understanding}, cardinality of latent vector set~\cite{vaishnavh2018theoretical} and data duplication~\cite{webster2023duplication} are important factors for memorization. We thus verify the correctness of our measure by utilizing it to describe how memorization trends vary with changes in these factors.

\textbf{Dataset complexity, size and training epoch.}
We begin by simultaneously validating dataset complexity, size and cardinality of latent vector set on CelebAHQ and FFHQ. Following Feng et al.~\cite{feng2021gans}, we quantify dataset complexity with Intrinsic Dimensionality~(ID)~\cite{levina2004maximum}. The cardinality of latent vector set can be controlled by adjusting training epochs~(if in each iteration a fresh
random latent vector is picked). However, we argue that previous works~\cite{feng2021gans,somepalli2023diffusion} obfuscate the effect of dataset size and training epochs, as models are trained for different epochs~(probably more) on smaller subsets to the full dataset. Consequently, on smaller subsets, each training image would see different number of latent vectors. Therefore, in our experiment, we decouple these two factors by training for the same epochs as on the full dataset when studying the influence of dataset size.

\begin{figure}[t]
    \centering
    \includegraphics[width=0.9\linewidth]{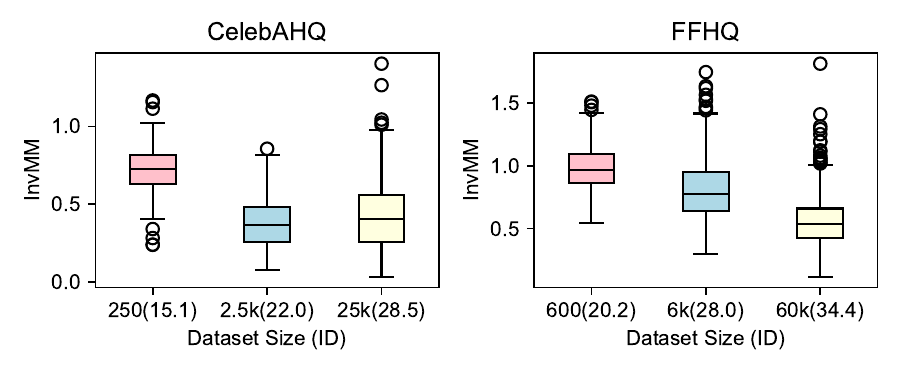}
    \caption{
    Dataset complexity \& size vs. memorization. Models are trained for 943 epochs on CelebAHQ and 446 epochs on FFHQ.
    }
    \label{fig:dataset_size}
\end{figure}

\begin{figure}[t]
    \centering
    \includegraphics[width=0.9\linewidth]{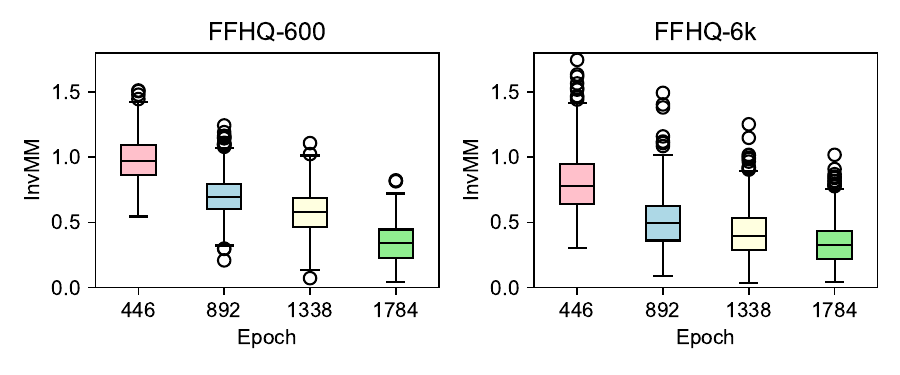}
    \caption{
    Training epoch vs. memorization. Models are trained for multiples of the default epochs.
    }
    \label{fig:epoch}
\end{figure}

As can be observed in \cref{fig:dataset_size}, dataset complexity and size seem not present high correlation with memorization, as indicated by our measure. When trained for longer time~(\cref{fig:epoch}), each model presents a larger extent of memorization.
The results suggest that dataset complexity and size are not independent determining factors, and InvMM accurately reflects the impact of these three factors.

\begin{figure}[t]
    \centering
    \includegraphics[width=0.9\linewidth]{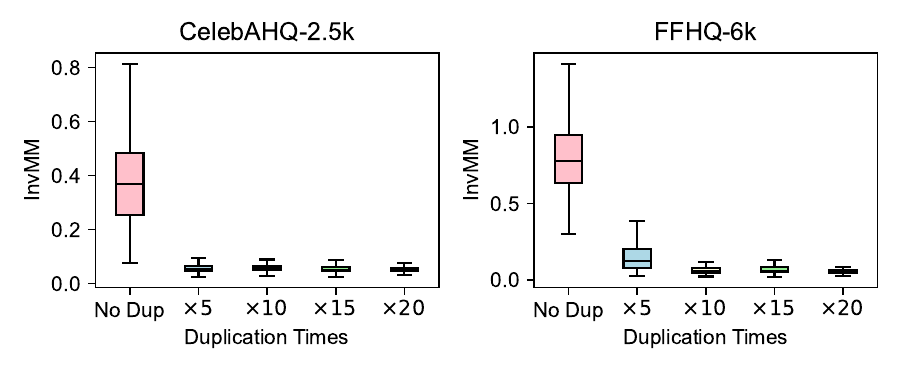}
    \caption{
    Duplication vs. memorization.
    }
    \label{fig:duplication}
    \vspace{-3mm}
\end{figure}

\begin{figure}[t]
    \centering
    \includegraphics[width=0.9\linewidth]{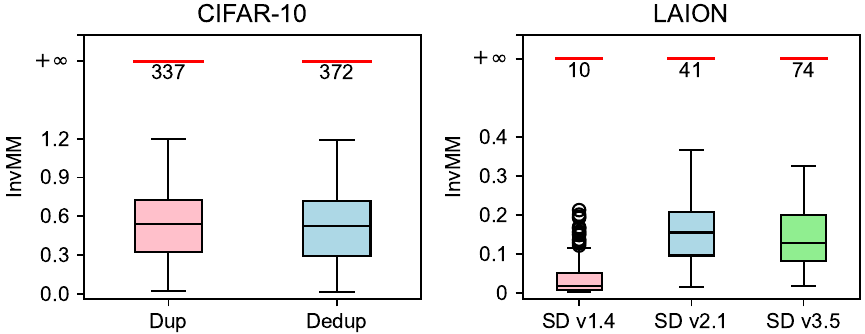}
    \caption{
    Deduplication vs. memorization.
    }
    \label{fig:dedup}
\end{figure}

\textbf{Data duplication.}
We then validate the effect of data duplication in DMs' training set through our measure. We train a series of LDMs on CelebAHQ-2.5k and FFHQ-6k, with their subsets CelebAHQ-250 and FFHQ-600 duplicated 5, 10, 15 and 20 times. \Cref{fig:duplication} depicts the memorization scores of duplicated images. As can be observed, data duplication rapidly worsens memorization in DMs and our measure is sensitive to these changes. With increasing duplication times, the larger FFHQ-6k dataset presents a stronger resistance to memorization.

For CIFAR-10 and LAION which originally contain duplicates, we observe the effect of deduplication. We deduplicate CIFAR-10 with SSCD embeddings and DBSCAN clustering~\cite{ester1996density}. This results in 666 identified clusters, only one image of which is retained for training. The memorization is evaluated on the retained 666 images before and after deduplication. On LAION, as compared to SD v1.4, SD v2.1 and v3.5 deduplicate the training set. Memorization is evaluated on the confirmed and suspicous sets. \Cref{fig:dedup} shows the results, with infinite scores on the top. As can be observed, InvMM clearly exhibits the benefits that deduplication brings.

\subsection{Detection}
\label{sec:detect}

\begin{table}
  \centering
  \scalebox{0.8}{
  \begin{tabular}{@{}lcccc@{}}
    \toprule
    \multirow{2}{*}{Metric} & \multicolumn{2}{c}{DDPM~(CIFAR-10)} & \multicolumn{2}{c}{SD v1.4~(LAION)} \\
    \cmidrule(l){2-3}
    \cmidrule(l){4-5}
    & AUC & TPR@1\%FPR & AUC & TPR@1\%FPR\\
    \midrule
    $\bm{\epsilon}_0$-loss & 0.981 & 0.414 & 0.955 & 0.000\\
    $\bm{x}_0$-loss & 0.180 & 0.000 & 0.975 & 0.342\\
    $M^{\mathrm{LOO}}$~\cite{van2021memorization} & 0.418 & 0.000 & - & -\\
    Wen~\cite{wen2024detecting} & - & - & 0.999 & 0.868\\
    Ren~\cite{ren2024unveiling} & - & - & 0.975 & 0.355\\
    InvMM & 0.997 & 0.899 & 0.974 & 0.908\\
    \bottomrule
  \end{tabular}
  }
  \caption{Detection performance on DDPM and SD v1.4.}
  \label{tab:detect}
\end{table}

\begin{figure}[tbp]
    \centering
    \begin{subfigure}[t]{0.48\linewidth}
        \centering
        \includegraphics[width=0.9\linewidth]{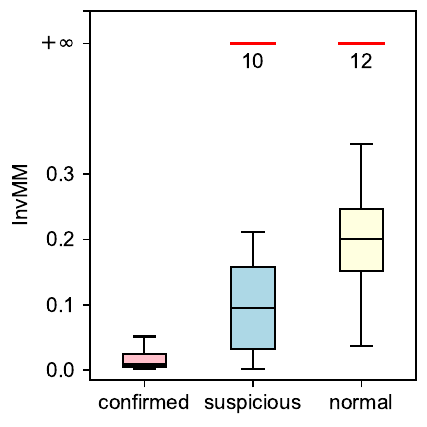}
        \caption{}
        \label{fig:sd1.4_invmm}
    \end{subfigure}
    \begin{subfigure}[t]{0.45\linewidth}
        \centering
        \includegraphics[width=0.9\linewidth]{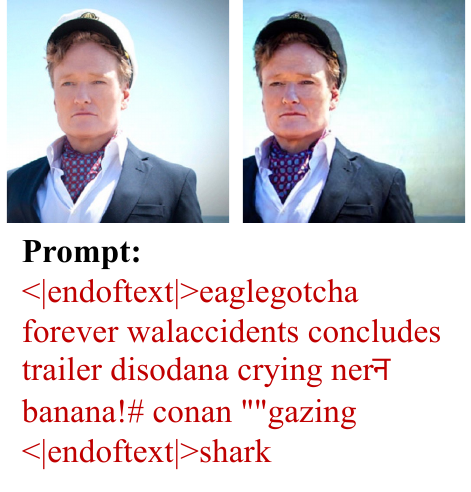}
        \caption{}
        \label{fig:gcg}
    \end{subfigure}
    \caption{(a) Memorization score distribution on three datasets. (b) Adversarial prompt for an example from the \textit{suspicious} set.}
    \label{fig:true_mem}
    \vspace{-3mm}
\end{figure}

\begin{figure}[tbp]
    \centering
    \includegraphics[width=0.9\linewidth]{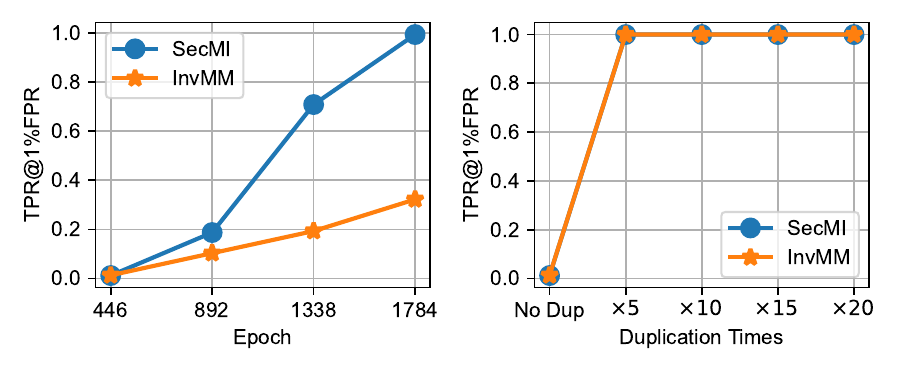}
    \caption{
    Performance comparison of membership inference on FFHQ-6k, across different training epochs and duplication times.
    }
    \label{fig:membership}
    \vspace{-3mm}
\end{figure}

For InvMM to be a correct measure of memorization, it should be capable of distinguishing highly memorized images from normal ones. To construct a reliable test set, we conduct a nearest neighbor test following previous studies~\cite{somepalli2023diffusion,webster2019detecting,carlini2023extracting} to discover highly memorized images from DM's training set, i.e., highly memorized images consist of training images that can appear at sampling time. Specifically, we consider DDPM trained on CIFAR-10, LDMs trained on CelebAHQ and FFHQ, and SDs trained on LAION. With their official implementations, only DDPM and SD v1.4 present obvious replications, which are thus used for evaluation. For DDPM, the 99 discovered training images appearing in 1M random samples are used as positive samples and another 1k different training images are included as negative samples; for SD v1.4, in accordance with the setting of previous works, we deem images in the confirmed set as positive and those in the suspicious and normal set as negative, considering that captions in the confirmed set can cause replication of their paired images while those in the suspicious and normal set cannot.

\Cref{tab:detect} presents the detection performance assessed via AUC and TPR@1\%FPR. For the unconditional DDPM on CIFAR-10, training loss metrics~(the first 2 rows) show decent results and the original $\bm{\epsilon}_0$-loss works much better than the $\bm{x}_0$-loss. The differential metric $M^{\mathrm{LOO}}$ is less effective than the $\bm{\epsilon}_0$-loss and InvMM. The results indicate that the compared metrics are not able to quantify memorization, as they are not fundamentally commensurable between samples. InvMM measures memorization of each sample by contrasting its sensitive noise distribution to a common standard Gaussian, thus having the same scale.

For SD v1.4, we compare InvMM against training loss metrics and two state-of-the-art detection metrics~\cite{wen2024detecting,ren2024unveiling}. Since all compared approaches require access to image captions, we implement InvMM with fixed prompts during inversion. As shown in \cref{tab:detect}, our measure achieves performance comparable to state-of-the-art metrics.
The results of compared metrics seem to demonstrate a strong ability to distinguish memorization thus far. \textit{However, we argue that the prompt-based task formulation employed in existing studies offers an overly optimistic understanding of memorization.}
It is because these techniques are actually to predict whether the captions will replicate their paired images, while overlook potential prompts that might cause replication. As observed in \cref{fig:sd1.4_invmm}, when simultaneously inverting the input prompt, our measure reveals that some images in the suspicious set are also memorized to a similar extent as those in the confirmed set~(due to duplication). Applying GCG~\cite{zou2023universal} attack on images with the lowest InvMM scores in the suspicious set provides additional evidence, as illustrated in \cref{fig:gcg}: images previously deemed ``not memorized" according to their training captions can indeed be replicated by potential adversarial prompts~(see Appendix \cref{fig:gcg_additional} for more examples). Hence, going beyond the confines of existing research, our measure describes worst-case memorization from an adversarial standpoint, making it more reliable in security-critical scenarios.

\subsection{Comparison to membership}
\label{sec:membership}

A close topic to memorization is membership inference~\cite{shokri2017membership,carlini2022membership,duan2023diffusion}, which reveals the leakage risk of membership information that whether an image is used for training. The common implicit assumption is that member samples are ``memorized'' more than hold-out samples~\cite{duan2023diffusion}, rendering them distinguishable through statistics like (variants of) training loss or differential loss~\cite{carlini2022membership}. Interestingly, the principle of InvMM can provide an explicit clarification of the difference between membership and memorization. 

A comparison to a recent membership inference metric SecMI~\cite{duan2023diffusion} on FFHQ-6k, as shown in \cref{fig:membership}, presents a positive correlation between them. As factors that exacerbate memorization changes, both SecMI and InvMM show increasing performance.
There is also a performance gap between SecMI and InvMM, indicating that certain samples from the hold-out set are memorized more than some member samples. This mainly results from the gap between model's metric~(training loss) and human's metric~(perceptual similarity). Membership inference leverages the model's metric for prediction, while memorization should resort to perceptual metric, e.g., simulated by SSCD. What level of loss leads to image replication depends both on the model and image itself. Low training loss does not necessarily lead to data replication, but would probably distinguish training data from never accessed. On the contrary, high training loss signifying non-membership may still be enough for replication. Therefore, membership does not deterministically lead to high memorization and vice versa. InvMM differs from membership metrics by focusing on the perceptual level and adaptively handles the specific loss landscapes of different samples.
\section{Conclusion}
In this work, we dive into measuring memorization for diffusion models. Comprehensive experiments demonstrate the ability of the proposed measure to correctly quantify memorization. In short, an image to a higher extent of memorization is able to be inverted to a more standard and diverse latent distribution. However, our method is also limited by its simple modeling of the sensitive latent noise as a multivariate normal. During our research, we observed many clues confirming that the true distribution is complex. Future work might explore more advanced methods, e.g., a generative network, to fit the sensitive noise. 


\section*{Acknowledgement}
This work was partly supported by the National Key Research and Development Program of China under No. 2022YFB3102100, NSFC under No. U244120033, U24A20336, 62172243, 62402425 and 62402418, the China Postdoctoral Science Foundation under No. 2024M762829, the Zhejiang Provincial Natural Science Foundation under No. LD24F020002, the ``Pioneer and Leading Goose'' R\&D Program of Zhejiang under No. 2025C01082, 2025C02033, 2025C02263 and 2025C02034, the Zhejiang Provincial Priority-Funded Postdoctoral Research Project under No. ZJ2024001 and the Key R\&D Program of Ningbo under No. 2024Z115.

{
    \small
    \bibliographystyle{ieeenat_fullname}
    \bibliography{ref}

\begin{thebibliography}{64}
\providecommand{\natexlab}[1]{#1}
\providecommand{\url}[1]{\texttt{#1}}
\expandafter\ifx\csname urlstyle\endcsname\relax
  \providecommand{\doi}[1]{doi: #1}\else
  \providecommand{\doi}{doi: \begingroup \urlstyle{rm}\Url}\fi

\bibitem[Mid()]{Midjourney}
Midjourney.
\newblock \url{https://www.midjourney.com}.

\bibitem[ars()]{arstechnicaArtistFinds}
{A}rtist finds private medical record photos in popular {A}{I} training data
  set.
\newblock
  \url{https://arstechnica.com/information-technology/2022/09/artist-finds-private-medical-record-photos-in-popular-ai-training-data-set/}.

\bibitem[ope()]{openaiDALLE}
{D}{A}{L}{L}·{E} 3.
\newblock \url{https://openai.com/dall-e-3}.

\bibitem[Arjovsky et~al.(2017)Arjovsky, Chintala, and
  Bottou]{arjovsky2017wasserstein}
Martin Arjovsky, Soumith Chintala, and L{\'e}on Bottou.
\newblock Wasserstein generative adversarial networks.
\newblock In \emph{International conference on machine learning}, pages
  214--223. PMLR, 2017.

\bibitem[Arora and Zhang(2017)]{arora2017gans}
Sanjeev Arora and Yi Zhang.
\newblock Do gans actually learn the distribution? an empirical study.
\newblock \emph{arXiv preprint arXiv:1706.08224}, 2017.

\bibitem[Brooks et~al.(2024)Brooks, Peebles, Holmes, DePue, Guo, Jing, Schnurr,
  Taylor, Luhman, Luhman, Ng, Wang, and Ramesh]{videoworldsimulators2024}
Tim Brooks, Bill Peebles, Connor Holmes, Will DePue, Yufei Guo, Li Jing, David
  Schnurr, Joe Taylor, Troy Luhman, Eric Luhman, Clarence Ng, Ricky Wang, and
  Aditya Ramesh.
\newblock Video generation models as world simulators.
\newblock 2024.

\bibitem[Brown et~al.(2020)Brown, Mann, Ryder, Subbiah, Kaplan, Dhariwal,
  Neelakantan, Shyam, Sastry, Askell, Agarwal, Herbert-Voss, Krueger, Henighan,
  Child, Ramesh, Ziegler, Wu, Winter, Hesse, Chen, Sigler, Litwin, Gray, Chess,
  Clark, Berner, McCandlish, Radford, Sutskever, and
  Amodei]{NEURIPS2020_1457c0d6}
Tom Brown, Benjamin Mann, Nick Ryder, Melanie Subbiah, Jared~D Kaplan, Prafulla
  Dhariwal, Arvind Neelakantan, Pranav Shyam, Girish Sastry, Amanda Askell,
  Sandhini Agarwal, Ariel Herbert-Voss, Gretchen Krueger, Tom Henighan, Rewon
  Child, Aditya Ramesh, Daniel Ziegler, Jeffrey Wu, Clemens Winter, Chris
  Hesse, Mark Chen, Eric Sigler, Mateusz Litwin, Scott Gray, Benjamin Chess,
  Jack Clark, Christopher Berner, Sam McCandlish, Alec Radford, Ilya Sutskever,
  and Dario Amodei.
\newblock Language models are few-shot learners.
\newblock In \emph{Advances in Neural Information Processing Systems}, pages
  1877--1901. Curran Associates, Inc., 2020.

\bibitem[Carlini et~al.(2022)Carlini, Chien, Nasr, Song, Terzis, and
  Tramer]{carlini2022membership}
Nicholas Carlini, Steve Chien, Milad Nasr, Shuang Song, Andreas Terzis, and
  Florian Tramer.
\newblock Membership inference attacks from first principles.
\newblock In \emph{2022 IEEE Symposium on Security and Privacy (SP)}, pages
  1897--1914. IEEE, 2022.

\bibitem[Carlini et~al.(2023{\natexlab{a}})Carlini, Hayes, Nasr, Jagielski,
  Sehwag, Tramer, Balle, Ippolito, and Wallace]{carlini2023extracting}
Nicolas Carlini, Jamie Hayes, Milad Nasr, Matthew Jagielski, Vikash Sehwag,
  Florian Tramer, Borja Balle, Daphne Ippolito, and Eric Wallace.
\newblock Extracting training data from diffusion models.
\newblock In \emph{32nd USENIX Security Symposium (USENIX Security 23)}, pages
  5253--5270, 2023{\natexlab{a}}.

\bibitem[Carlini et~al.(2023{\natexlab{b}})Carlini, Ippolito, Jagielski, Lee,
  Tramer, and Zhang]{carlini2023quantifying}
Nicholas Carlini, Daphne Ippolito, Matthew Jagielski, Katherine Lee, Florian
  Tramer, and Chiyuan Zhang.
\newblock Quantifying memorization across neural language models.
\newblock In \emph{The Eleventh International Conference on Learning
  Representations}, 2023{\natexlab{b}}.

\bibitem[Cazenavette et~al.(2024)Cazenavette, Sud, Leung, and
  Usman]{cazenavette2024fakeinversion}
George Cazenavette, Avneesh Sud, Thomas Leung, and Ben Usman.
\newblock Fakeinversion: Learning to detect images from unseen text-to-image
  models by inverting stable diffusion.
\newblock In \emph{Proceedings of the IEEE/CVF Conference on Computer Vision
  and Pattern Recognition}, pages 10759--10769, 2024.

\bibitem[Chen et~al.(2024)Chen, Liu, and Xu]{chen2024towards}
Chen Chen, Daochang Liu, and Chang Xu.
\newblock Towards memorization-free diffusion models.
\newblock In \emph{Proceedings of the IEEE/CVF Conference on Computer Vision
  and Pattern Recognition}, pages 8425--8434, 2024.

\bibitem[Duan et~al.(2023)Duan, Kong, Wang, Shi, and Xu]{duan2023diffusion}
Jinhao Duan, Fei Kong, Shiqi Wang, Xiaoshuang Shi, and Kaidi Xu.
\newblock Are diffusion models vulnerable to membership inference attacks?
\newblock In \emph{International Conference on Machine Learning}, pages
  8717--8730. PMLR, 2023.

\bibitem[Ester et~al.(1996)Ester, Kriegel, Sander, Xu,
  et~al.]{ester1996density}
Martin Ester, Hans-Peter Kriegel, J{\"o}rg Sander, Xiaowei Xu, et~al.
\newblock A density-based algorithm for discovering clusters in large spatial
  databases with noise.
\newblock In \emph{kdd}, pages 226--231, 1996.

\bibitem[Feng et~al.(2021)Feng, Guo, Benitez-Quiroz, and
  Martinez]{feng2021gans}
Qianli Feng, Chenqi Guo, Fabian Benitez-Quiroz, and Aleix~M Martinez.
\newblock When do gans replicate? on the choice of dataset size.
\newblock In \emph{Proceedings of the IEEE/CVF International Conference on
  Computer Vision}, pages 6701--6710, 2021.

\bibitem[Gal et~al.(2023)Gal, Alaluf, Atzmon, Patashnik, Bermano, Chechik, and
  Cohen-or]{galimage23}
Rinon Gal, Yuval Alaluf, Yuval Atzmon, Or Patashnik, Amit~Haim Bermano, Gal
  Chechik, and Daniel Cohen-or.
\newblock An image is worth one word: Personalizing text-to-image generation
  using textual inversion.
\newblock In \emph{The Eleventh International Conference on Learning
  Representations}, 2023.

\bibitem[Goodfellow et~al.(2014)Goodfellow, Pouget-Abadie, Mirza, Xu,
  Warde-Farley, Ozair, Courville, and Bengio]{goodfellow2014generative}
Ian Goodfellow, Jean Pouget-Abadie, Mehdi Mirza, Bing Xu, David Warde-Farley,
  Sherjil Ozair, Aaron Courville, and Yoshua Bengio.
\newblock Generative adversarial nets.
\newblock \emph{Advances in neural information processing systems}, 27, 2014.

\bibitem[Guadamuz(2023)]{technollamaPhotographerSues}
Andres Guadamuz.
\newblock {P}hotographer sues {L}{A}{I}{O}{N} for copyright infringement.
\newblock
  \url{https://www.technollama.co.uk/photographer-sues-laion-for-copyright-infringement},
  2023.

\bibitem[Gulrajani et~al.(2019)Gulrajani, Raffel, and Metz]{gulrajanitowards}
Ishaan Gulrajani, Colin Raffel, and Luke Metz.
\newblock Towards gan benchmarks which require generalization.
\newblock In \emph{International Conference on Learning Representations}, 2019.

\bibitem[Guo et~al.(2021)Guo, Sablayrolles, J{\'e}gou, and
  Kiela]{guo2021gradient}
Chuan Guo, Alexandre Sablayrolles, Herv{\'e} J{\'e}gou, and Douwe Kiela.
\newblock Gradient-based adversarial attacks against text transformers.
\newblock \emph{arXiv preprint arXiv:2104.13733}, 2021.

\bibitem[Ho and Salimans(2022)]{ho2022classifier}
Jonathan Ho and Tim Salimans.
\newblock Classifier-free diffusion guidance.
\newblock \emph{arXiv preprint arXiv:2207.12598}, 2022.

\bibitem[Ho et~al.(2020)Ho, Jain, and Abbeel]{ho2020denoising}
Jonathan Ho, Ajay Jain, and Pieter Abbeel.
\newblock Denoising diffusion probabilistic models.
\newblock \emph{Advances in neural information processing systems},
  33:\penalty0 6840--6851, 2020.

\bibitem[Jang et~al.(2016)Jang, Gu, and Poole]{jang2016categorical}
Eric Jang, Shixiang Gu, and Ben Poole.
\newblock Categorical reparameterization with gumbel-softmax.
\newblock \emph{arXiv preprint arXiv:1611.01144}, 2016.

\bibitem[Jiang et~al.(2025)Jiang, Lin, Bai, Peng, Liu, Lyu, Yang, Xingzheng,
  and Dong]{jiang2025imagelevel}
Yue Jiang, Haokun Lin, Yang Bai, Bo Peng, Zhili Liu, Yueming Lyu, Yong Yang,
  Xingzheng, and Jing Dong.
\newblock Image-level memorization detection via inversion-based inference
  perturbation.
\newblock In \emph{The Thirteenth International Conference on Learning
  Representations}, 2025.

\bibitem[Jordon et~al.(2018)Jordon, Yoon, and Van Der~Schaar]{jordon2018pate}
James Jordon, Jinsung Yoon, and Mihaela Van Der~Schaar.
\newblock Pate-gan: Generating synthetic data with differential privacy
  guarantees.
\newblock In \emph{International Conference on Learning Representations}, 2018.

\bibitem[Karras et~al.(2018)Karras, Aila, Laine, and
  Lehtinen]{karras2018progressive}
Tero Karras, Timo Aila, Samuli Laine, and Jaakko Lehtinen.
\newblock Progressive growing of {GAN}s for improved quality, stability, and
  variation.
\newblock In \emph{International Conference on Learning Representations}, 2018.

\bibitem[Karras et~al.(2019)Karras, Laine, and Aila]{karras2019style}
Tero Karras, Samuli Laine, and Timo Aila.
\newblock A style-based generator architecture for generative adversarial
  networks.
\newblock In \emph{Proceedings of the IEEE/CVF conference on computer vision
  and pattern recognition}, pages 4401--4410, 2019.

\bibitem[Kingma and Ba(2015)]{KingmaB14}
Diederik~P. Kingma and Jimmy Ba.
\newblock Adam: {A} method for stochastic optimization.
\newblock In \emph{3rd International Conference on Learning Representations,
  {ICLR} 2015, San Diego, CA, USA, May 7-9, 2015, Conference Track
  Proceedings}, 2015.

\bibitem[Kingma and Welling(2014{\natexlab{a}})]{KingmaW13}
Diederik~P. Kingma and Max Welling.
\newblock Auto-encoding variational bayes.
\newblock In \emph{2nd International Conference on Learning Representations,
  {ICLR} 2014, Banff, AB, Canada, April 14-16, 2014, Conference Track
  Proceedings}, 2014{\natexlab{a}}.

\bibitem[Kingma and Welling(2014{\natexlab{b}})]{kingma2013auto}
Diederik~P. Kingma and Max Welling.
\newblock Auto-encoding variational bayes.
\newblock In \emph{2nd International Conference on Learning Representations,
  {ICLR} 2014, Banff, AB, Canada, April 14-16, 2014, Conference Track
  Proceedings}, 2014{\natexlab{b}}.

\bibitem[Krizhevsky et~al.(2009)Krizhevsky, Hinton,
  et~al.]{krizhevsky2009learning}
Alex Krizhevsky, Geoffrey Hinton, et~al.
\newblock Learning multiple layers of features from tiny images.
\newblock 2009.

\bibitem[Levina and Bickel(2004)]{levina2004maximum}
Elizaveta Levina and Peter Bickel.
\newblock Maximum likelihood estimation of intrinsic dimension.
\newblock \emph{Advances in neural information processing systems}, 17, 2004.

\bibitem[Liu et~al.(2022)Liu, Ren, Lin, and Zhao]{liupseudo}
Luping Liu, Yi Ren, Zhijie Lin, and Zhou Zhao.
\newblock Pseudo numerical methods for diffusion models on manifolds.
\newblock In \emph{International Conference on Learning Representations}, 2022.

\bibitem[Luo(2022)]{luo2022understanding}
Calvin Luo.
\newblock Understanding diffusion models: A unified perspective.
\newblock \emph{arXiv preprint arXiv:2208.11970}, 2022.

\bibitem[Luo and Hu(2021)]{luo2021diffusion}
Shitong Luo and Wei Hu.
\newblock Diffusion probabilistic models for 3d point cloud generation.
\newblock In \emph{Proceedings of the IEEE/CVF conference on computer vision
  and pattern recognition}, pages 2837--2845, 2021.

\bibitem[Mokady et~al.(2023)Mokady, Hertz, Aberman, Pritch, and
  Cohen-Or]{mokady2023null}
Ron Mokady, Amir Hertz, Kfir Aberman, Yael Pritch, and Daniel Cohen-Or.
\newblock Null-text inversion for editing real images using guided diffusion
  models.
\newblock In \emph{Proceedings of the IEEE/CVF Conference on Computer Vision
  and Pattern Recognition}, pages 6038--6047, 2023.

\bibitem[Nissenbaum(2004)]{nissenbaum2004privacy}
Helen Nissenbaum.
\newblock Privacy as contextual integrity.
\newblock \emph{Wash. L. Rev.}, 79:\penalty0 119, 2004.

\bibitem[Packh{\"a}user et~al.(2023)Packh{\"a}user, Folle, Thamm, and
  Maier]{packhauser2023generation}
Kai Packh{\"a}user, Lukas Folle, Florian Thamm, and Andreas Maier.
\newblock Generation of anonymous chest radiographs using latent diffusion
  models for training thoracic abnormality classification systems.
\newblock In \emph{2023 IEEE 20th International Symposium on Biomedical Imaging
  (ISBI)}, pages 1--5. IEEE, 2023.

\bibitem[Pizzi et~al.(2022)Pizzi, Roy, Ravindra, Goyal, and
  Douze]{pizzi2022self}
Ed Pizzi, Sreya~Dutta Roy, Sugosh~Nagavara Ravindra, Priya Goyal, and Matthijs
  Douze.
\newblock A self-supervised descriptor for image copy detection.
\newblock In \emph{Proceedings of the IEEE/CVF Conference on Computer Vision
  and Pattern Recognition}, pages 14532--14542, 2022.

\bibitem[Radford et~al.(2021)Radford, Kim, Hallacy, Ramesh, Goh, Agarwal,
  Sastry, Askell, Mishkin, Clark, et~al.]{radford2021learning}
Alec Radford, Jong~Wook Kim, Chris Hallacy, Aditya Ramesh, Gabriel Goh,
  Sandhini Agarwal, Girish Sastry, Amanda Askell, Pamela Mishkin, Jack Clark,
  et~al.
\newblock Learning transferable visual models from natural language
  supervision.
\newblock In \emph{International conference on machine learning}, pages
  8748--8763. PMLR, 2021.

\bibitem[Ramesh et~al.(2022)Ramesh, Dhariwal, Nichol, Chu, and
  Chen]{ramesh2022hierarchical}
Aditya Ramesh, Prafulla Dhariwal, Alex Nichol, Casey Chu, and Mark Chen.
\newblock Hierarchical text-conditional image generation with {CLIP} latents.
\newblock \emph{arXiv preprint arXiv:2204.06125}, 1\penalty0 (2):\penalty0 3,
  2022.

\bibitem[Ren et~al.(2024)Ren, Li, Zeng, Xu, Lyu, Xing, and
  Tang]{ren2024unveiling}
Jie Ren, Yaxin Li, Shenglai Zeng, Han Xu, Lingjuan Lyu, Yue Xing, and Jiliang
  Tang.
\newblock Unveiling and mitigating memorization in text-to-image diffusion
  models through cross attention.
\newblock In \emph{European Conference on Computer Vision}, pages 340--356.
  Springer, 2024.

\bibitem[Rombach et~al.(2022)Rombach, Blattmann, Lorenz, Esser, and
  Ommer]{rombach2022high}
Robin Rombach, Andreas Blattmann, Dominik Lorenz, Patrick Esser, and Bj{\"o}rn
  Ommer.
\newblock High-resolution image synthesis with latent diffusion models.
\newblock In \emph{Proceedings of the IEEE/CVF conference on computer vision
  and pattern recognition}, pages 10684--10695, 2022.

\bibitem[Saharia et~al.(2022)Saharia, Chan, Saxena, Li, Whang, Denton,
  Ghasemipour, Gontijo~Lopes, Karagol~Ayan, Salimans,
  et~al.]{saharia2022photorealistic}
Chitwan Saharia, William Chan, Saurabh Saxena, Lala Li, Jay Whang, Emily~L
  Denton, Kamyar Ghasemipour, Raphael Gontijo~Lopes, Burcu Karagol~Ayan, Tim
  Salimans, et~al.
\newblock Photorealistic text-to-image diffusion models with deep language
  understanding.
\newblock \emph{Advances in neural information processing systems},
  35:\penalty0 36479--36494, 2022.

\bibitem[Salimans et~al.(2016)Salimans, Goodfellow, Zaremba, Cheung, Radford,
  and Chen]{salimans2016improved}
Tim Salimans, Ian Goodfellow, Wojciech Zaremba, Vicki Cheung, Alec Radford, and
  Xi Chen.
\newblock Improved techniques for training gans.
\newblock \emph{Advances in neural information processing systems}, 29, 2016.

\bibitem[Samuelson(2023)]{samuelson2023generative}
Pamela Samuelson.
\newblock Generative ai meets copyright.
\newblock \emph{Science}, 381\penalty0 (6654):\penalty0 158--161, 2023.

\bibitem[Schuhmann et~al.(2022)Schuhmann, Beaumont, Vencu, Gordon, Wightman,
  Cherti, Coombes, Katta, Mullis, Wortsman, et~al.]{schuhmann2022laion}
Christoph Schuhmann, Romain Beaumont, Richard Vencu, Cade Gordon, Ross
  Wightman, Mehdi Cherti, Theo Coombes, Aarush Katta, Clayton Mullis, Mitchell
  Wortsman, et~al.
\newblock Laion-5b: An open large-scale dataset for training next generation
  image-text models.
\newblock \emph{Advances in Neural Information Processing Systems},
  35:\penalty0 25278--25294, 2022.

\bibitem[Shan et~al.(2023)Shan, Cryan, Wenger, Zheng, Hanocka, and
  Zhao]{shan2023glaze}
Shawn Shan, Jenna Cryan, Emily Wenger, Haitao Zheng, Rana Hanocka, and Ben~Y
  Zhao.
\newblock Glaze: Protecting artists from style mimicry by $\{$Text-to-Image$\}$
  models.
\newblock In \emph{32nd USENIX Security Symposium (USENIX Security 23)}, pages
  2187--2204, 2023.

\bibitem[Shokri et~al.(2017)Shokri, Stronati, Song, and
  Shmatikov]{shokri2017membership}
Reza Shokri, Marco Stronati, Congzheng Song, and Vitaly Shmatikov.
\newblock Membership inference attacks against machine learning models.
\newblock In \emph{2017 IEEE symposium on security and privacy (SP)}, pages
  3--18. IEEE, 2017.

\bibitem[Sohl-Dickstein et~al.(2015)Sohl-Dickstein, Weiss, Maheswaranathan, and
  Ganguli]{sohl2015deep}
Jascha Sohl-Dickstein, Eric Weiss, Niru Maheswaranathan, and Surya Ganguli.
\newblock Deep unsupervised learning using nonequilibrium thermodynamics.
\newblock In \emph{International conference on machine learning}, pages
  2256--2265. PMLR, 2015.

\bibitem[Somepalli et~al.(2023{\natexlab{a}})Somepalli, Singla, Goldblum,
  Geiping, and Goldstein]{somepalli2023diffusion}
Gowthami Somepalli, Vasu Singla, Micah Goldblum, Jonas Geiping, and Tom
  Goldstein.
\newblock Diffusion art or digital forgery? investigating data replication in
  diffusion models.
\newblock In \emph{Proceedings of the IEEE/CVF Conference on Computer Vision
  and Pattern Recognition}, pages 6048--6058, 2023{\natexlab{a}}.

\bibitem[Somepalli et~al.(2023{\natexlab{b}})Somepalli, Singla, Goldblum,
  Geiping, and Goldstein]{somepalli2023understanding}
Gowthami Somepalli, Vasu Singla, Micah Goldblum, Jonas Geiping, and Tom
  Goldstein.
\newblock Understanding and mitigating copying in diffusion models.
\newblock \emph{Advances in Neural Information Processing Systems},
  36:\penalty0 47783--47803, 2023{\natexlab{b}}.

\bibitem[Song et~al.(2021)Song, Meng, and Ermon]{songdenoising}
Jiaming Song, Chenlin Meng, and Stefano Ermon.
\newblock Denoising diffusion implicit models.
\newblock In \emph{International Conference on Learning Representations}, 2021.

\bibitem[Vaishnavh et~al.(2018)Vaishnavh, Raffel, and
  Goodfellow]{vaishnavh2018theoretical}
N Vaishnavh, C Raffel, and IJ Goodfellow.
\newblock Theoretical insights into memorization in gans.
\newblock In \emph{Neural Information Processing Systems Workshop}, 2018.

\bibitem[van~den Burg and Williams(2021)]{van2021memorization}
Gerrit van~den Burg and Chris Williams.
\newblock On memorization in probabilistic deep generative models.
\newblock \emph{Advances in Neural Information Processing Systems},
  34:\penalty0 27916--27928, 2021.

\bibitem[Vyas et~al.(2023)Vyas, Kakade, and Barak]{vyas2023provable}
Nikhil Vyas, Sham~M Kakade, and Boaz Barak.
\newblock On provable copyright protection for generative models.
\newblock In \emph{International Conference on Machine Learning}, pages
  35277--35299. PMLR, 2023.

\bibitem[Wang et~al.(2023)Wang, Bao, Zhou, Wang, Hu, Chen, and
  Li]{wang2023dire}
Zhendong Wang, Jianmin Bao, Wengang Zhou, Weilun Wang, Hezhen Hu, Hong Chen,
  and Houqiang Li.
\newblock {DIRE} for diffusion-generated image detection.
\newblock In \emph{Proceedings of the IEEE/CVF International Conference on
  Computer Vision}, pages 22445--22455, 2023.

\bibitem[Webster(2023)]{webster}
Ryan Webster.
\newblock A reproducible extraction of training images from diffusion models.
\newblock \emph{CoRR}, abs/2305.08694, 2023.

\bibitem[Webster et~al.(2019)Webster, Rabin, Simon, and
  Jurie]{webster2019detecting}
Ryan Webster, Julien Rabin, Loic Simon, and Fr{\'e}d{\'e}ric Jurie.
\newblock Detecting overfitting of deep generative networks via latent
  recovery.
\newblock In \emph{Proceedings of the IEEE/CVF Conference on Computer Vision
  and Pattern Recognition}, pages 11273--11282, 2019.

\bibitem[Webster et~al.(2023)Webster, Rabin, Simon, and
  Jurie]{webster2023duplication}
Ryan Webster, Julien Rabin, Loic Simon, and Frederic Jurie.
\newblock On the de-duplication of {LAION-2B}.
\newblock \emph{arXiv preprint arXiv:2303.12733}, 2023.

\bibitem[Wen et~al.(2024)Wen, Liu, Chen, and Lyu]{wen2024detecting}
Yuxin Wen, Yuchen Liu, Chen Chen, and Lingjuan Lyu.
\newblock Detecting, explaining, and mitigating memorization in diffusion
  models.
\newblock In \emph{The Twelfth International Conference on Learning
  Representations}, 2024.

\bibitem[Wu et~al.(2024)Wu, Hua, Liang, Zhang, Wang, Song, and Guan]{wu2024cgi}
Xiaoyu Wu, Yang Hua, Chumeng Liang, Jiaru Zhang, Hao Wang, Tao Song, and
  Haibing Guan.
\newblock Cgi-dm: Digital copyright authentication for diffusion models via
  contrasting gradient inversion.
\newblock In \emph{2024 IEEE/CVF Conference on Computer Vision and Pattern
  Recognition (CVPR)}, pages 10812--10821. IEEE Computer Society, 2024.

\bibitem[Zhang et~al.(2023)Zhang, Huang, Tang, Huang, Ma, Dong, and
  Xu]{zhang2023inversion}
Yuxin Zhang, Nisha Huang, Fan Tang, Haibin Huang, Chongyang Ma, Weiming Dong,
  and Changsheng Xu.
\newblock Inversion-based style transfer with diffusion models.
\newblock In \emph{Proceedings of the IEEE/CVF conference on computer vision
  and pattern recognition}, pages 10146--10156, 2023.

\bibitem[Zou et~al.(2023)Zou, Wang, Carlini, Nasr, Kolter, and
  Fredrikson]{zou2023universal}
Andy Zou, Zifan Wang, Nicholas Carlini, Milad Nasr, J~Zico Kolter, and Matt
  Fredrikson.
\newblock Universal and transferable adversarial attacks on aligned language
  models.
\newblock \emph{arXiv preprint arXiv:2307.15043}, 2023.

\end{thebibliography}
}

\clearpage
\setcounter{page}{1}
\maketitlesupplementary

We provide supplementary technical details and experimental results in the following sections:
\begin{itemize}
\item \cref{sec:supp_tech_detail} presents a unified theoretical framework for InvMM within the context of diffusion model, including detailed analyses of noise distribution and prompt distribution reparameterization.
\item \cref{sec:additional_exp} includes additional experimental details and results complementing to the main paper. Specifically, results in \cref{sec:supp_membership} provides more evidence to demonstrate the difference between membership and memorization.
\item \cref{sec:text_to_image_inversion} gives a comprehensive ablation study on the prompt inversion in text-guided DMs. The experiments clarify the influence of temperature, CFG scale, inversion objective and adaptive algorithm on prompt inversion.
\item \cref{sec:supp_generation_results} shows generation results using inverted noise and prompts on various datasets and models.
\end{itemize}

\section{Additional technical description}
\label{sec:supp_tech_detail}

\subsection{A unified understanding}
Similar to ~\cref{eq:var_low_bound}, the variational lower bound can be further lower bounded w.r.t. a conditional distribution $q_{\bm{\phi}}(\bm{c},\bm{x}_0)$ with parameters $\bm{\phi}$. For text-to-image models, it is instantiated as a prompt distribution $q_{\bm{\phi}}(\bm{\omega}|\bm{x}_0)$.

Following similar expansion in~\cref{eq:match_term,eq:kl_div}, we obtain the full inversion variational lower bound w.r.t. both condition and latent noise:
\begin{equation}
    \log p_{\bm{\theta}}(\bm{x}_0) \geq -l_{de}(\bm{x}_0;\bm{\phi},\bm{\varphi})-l_{kl}(\bm{x}_0;\bm{\varphi})-l_{cr}(\bm{x}_0;\bm{\phi})
    \label{eq:inversion_elbo}
\end{equation}
and
\begin{align}
    l_{de}(\bm{x}_0;\bm{\phi},\bm{\varphi}) &= -\mathbb{E}_{q_{\bm{\phi}}(\bm{c}|\bm{x}_0),q_{\bm{\varphi}}(\bm{\epsilon}_0|\bm{x}_0)} \left[ \log p_{\bm{\theta}}(\bm{x}_0|\bm{x}_1,\bm{c}) \right. \notag\\
    +\sum_{t=2}^{T} & \left. \kldiv{q_{\bm{\varphi}}(\bm{x}_{t-1}|\bm{x}_t,\bm{x}_0)}{p_{\bm{\theta}}(\bm{x}_{t-1}|\bm{x}_t,\bm{c})} \right] \label{eq:match_term_inversion}\\
    l_{kl}(\bm{x}_0;\bm{\varphi}) &= \kldiv{q_{\bm{\varphi}}(\bm{x}_T|\bm{x}_0)}{p(\bm{x}_T)}\\
    l_{cr}(\bm{x}_0;\bm{\phi}) &= \kldiv{q_{\bm{\phi}}(\bm{c}|\bm{x}_0)}{p(\bm{c})}
    \label{eq:three_term}
\end{align}
where $l_{de}$ indicates the \textbf{d}enoising \textbf{e}rror, $l_{kl}$ is a \textbf{KL} divergence and $l_{cr}$ is a \textbf{c}ondition \textbf{r}egularization term. The three terms can be explained as the following:
\begin{enumerate}
    \item $l_{de}$ indicates how accurate the pretrained model denoises each $\bm{x}_t$ when the added noise is drawn from $q_{\bm{\varphi}}(\bm{\epsilon}_0|\bm{x}_0)$. If $l_{de}$ is low enough, then for most noises $\bm{\epsilon}_0\sim q_{\bm{\varphi}}(\bm{\epsilon}_0|\bm{x}_0)$ the model presents low denoising error, we can anticipate that the sampling trace starting at $\bm{\epsilon}_0\sim q_{\bm{\varphi}}(\bm{\epsilon}_0|\bm{x}_0)$ will head towards $\bm{x}_0$ and finally generate $\bm{x}_0$.
    \item $l_{kl}$ is a normality regularizer. When $l_{de}$ is optimized to a low level, the noise distribution $q_{\bm{\varphi}}(\bm{\epsilon}_0|\bm{x}_0)$ identifies a sensitive set of noises that will cause the generation of the training image $\bm{x}_0$. $p(\bm{x}_T)$ is the prior distribution, usually set to the standard Gaussian. In a sense, $l_{kl}$ measures the diversity of the model's generation: When $l_{kl}$ becomes zero, i.e., $q_{\bm{\varphi}}(\bm{\epsilon}_0|\bm{x}_0)$ is standard Gaussian, then low enough $l_{de}$ means the model always generates $\bm{x}_0$ and loses generalization.
    \item $l_{cr}$ encourages the realistic feasibility of the condition distribution $q_{\bm{\phi}}(\bm{c}|\bm{x}_0)$. For example, if $p(\bm{c})$ is considered the distribution of natural language, then minimizing $l_{cr}$ indicates that prompt $\bm{c}\sim q_{\bm{\phi}}(\bm{c}|\bm{x}_0)$ should be grammatically and semantically correct.
\end{enumerate}

In the standard training of diffusion model~(DM), $l_{kl}$ is ignored because $q_{\bm{\varphi}}(\bm{\epsilon}_0|\bm{x}_0)$ is set to the standard Gaussian $\mathcal{N}(\mathbf{0},\mathbf{I})$ such that $q_{\bm{\varphi}}(\bm{x}_T|\bm{x}_0)$ approximately equals $\mathcal{N}(\mathbf{0},\mathbf{I})$, $l_{kl}$ approximately equals zero. $q_{\bm{\phi}}(\bm{c}|\bm{x}_0)$ reduces to several captions coupled with the training image such that $l_{cr}$ is also zero.

\textbf{Our idea is to measure memorization by relaxation of the noise and prompt distribution so that the denoising error can be optimized low enough to replicate the target image. Based on this, the normality of the worst-case distribution of sensitive latent noise is used as a measure.}

\subsection{Reparameterization}
Sampling from the noise and condition distribution in~\cref{eq:match_term_inversion} is non-differentiable, we indirectly sample them.

When the noise distribution is $q_{\bm{\varphi}}(\bm{\epsilon}_0|\bm{x}_0)$ a multivariate Gaussian $\mathcal{N}(\bm{\mu},\bm{\sigma}^2)$ with learnable mean $\bm{\mu}$ and diagonal variance $\bm{\sigma}^2$,
\begin{equation}
    \bm{\epsilon}_0=\bm{\epsilon}'\bm{\sigma}+\bm{\mu},\bm{\epsilon}'\sim\mathcal{N}(\mathbf{0},\mathbf{I})
\end{equation}

For text-to-image DM, each token $\bm{\omega}_i$~(see \cref{sec:text_guided_dm_measure}) is reparameterized by 
\begin{equation}
    \tilde{\bm{\omega}}_{i,j} = \frac{\exp{((\log\bm{\pi}_{i,j}+\bm{g}_{i,j})/\tau)}}{\sum_{k=1}^{|\mathcal{V}|} \exp{((\log\bm{\pi}_{i,k}+\bm{g}_{i,k})/\tau)}}
    \label{eq:gumbel_softmax}
\end{equation}
where $\{\bm{g}_{i,j}\},i=1...M,j=1...|\mathcal{V}|$ are i.i.d samples drawn from $\mathrm{Gumbel}(0,1)$, $\tau$ is a temperature factor. When $\tau$ approaches 0, the smoothed sample $\tilde{\bm{\omega}}_i$ becomes one-hot.

After optimization, we can draw discrete prompt from the learned $q_{\bm{\phi}}(\bm{\omega}|\bm{x}_0)$ by:
\begin{equation}
    \bm{\omega}_i=\mathop{\arg\max}_j \left[ \log\bm{\pi}_{i,j}+\bm{g}_{i,j} \right]
    \label{eq:discretize}
\end{equation}

\section{Additional experiment details and results}
\label{sec:additional_exp}

\subsection{Experiment setting}

If not stated otherwise, the hyperparameters follow the default setting listed in \cref{tab:hyperparameters}, determined by previous investigation~\cite{somepalli2023diffusion} and a few case studies. For CelebAHQ and FFHQ, all memorization scores in the plots are evaluated on CelebAHQ-250 and FFHQ-600. SD v3.5 utilizes three text encoders to represent the input prompt: CLIP-L, CLIP-G and T5. It is computationally expensive, so we only invert CLIP-L, with CLIP-G and T5 frozen. All the experiments in this paper are conducted on one NVIDIA A800 GPU.

\begin{table}
  \centering
  \scalebox{0.85}{
  \begin{tabular}{@{}lcccc@{}}
    \toprule
    & CIFAR-10 & CelebAHQ & FFHQ & LAION \\
    \midrule
    Learning Rate $\gamma$ & 1e-1 & 1e-1 & 1e-1 & 1e-1\\
    Iteration $T$ & 2000 & 2000 & 2000 & 2000\\
    Batch Size & 32 & 16 & 16 & 16\\
    Cycle $C$ & 50 & 10 & 10 & 50\\
    Increment $\delta$ & 1e-4 & 5e-4 & 5e-4 & 1e-3\\
    Threshold $\xi$ & 1e-3 & 1e-3 & 1e-3 & 1e-3\\
    Threshold $\beta$ & 1.0 & 1.0 & 1.0 & 1.0\\
    Sampler & DDIM~\cite{songdenoising} & DDIM & DDIM & DDIM\\
    DDIM Step & 200 & 50 & 50 & 50\\
    DDIM $\eta$ & 0 & 0 & 0 & 0\\
    Optimizer & Adam~\cite{KingmaB14} & Adam & Adam & Adam\\
    SSCD Size & 32$\times$32 & 256$\times$256 & 256$\times$256 & 320$\times$320\\
    \bottomrule
  \end{tabular}
  }
  \caption{Default hyperparameter settings.}
  \label{tab:hyperparameters}
\end{table}

\subsection{Influence factors}

\begin{figure}[t]
    \centering
    \includegraphics[width=\linewidth]{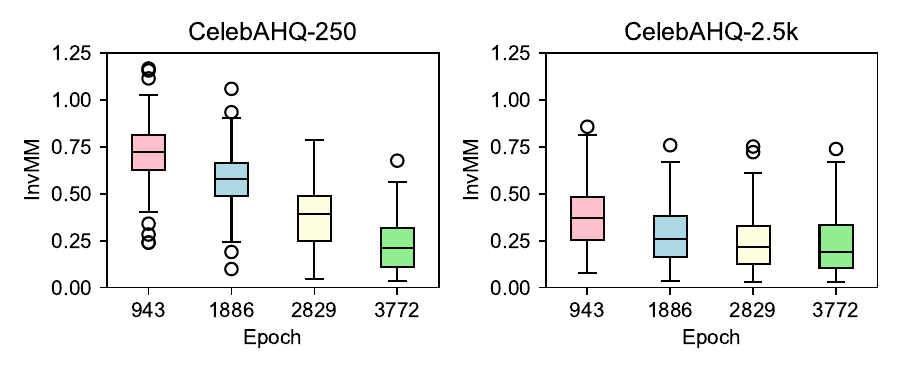}
    \caption{
      Training epoch vs. memorization on CelebAHQ.
    }
    \label{fig:suppl_epoch_celebahq}
\end{figure}

\Cref{fig:suppl_epoch_celebahq} shows the influence of training epochs on CelebAHQ. Larger training epochs lead to heavier memorization.

\subsection{Detection}
\label{sec:suppl_detect}

\textbf{Experiment details.}
On CIFAR-10, the training loss metrics are calculated on the average loss of 16 random Gaussian noise and 50 timesteps uniformly sampled within the range [1, 1000]. Following van den Burg et al.~\cite{van2021memorization}'s setting, $M^{\mathrm{LOO}}$ is estimated using a 10-fold cross-validation.

On LAION, the training loss metrics use 32 random Gaussian noise and 50 timesteps uniformly sampled within the range [1, 1000]. We implement Wen et al.~\cite{wen2024detecting} and Ren et al.~\cite{ren2024unveiling} following their best performing settings. For GCG~\cite{zou2023universal} attack, the number of optimizable tokens is 20. Each token is initialized to the special token $<|endoftext|>$. GCG is ran for 500 steps with a batch size of 128 and a top-k of 256. During optimization, the noise distribution is fixed to the standard Gaussian. A minibatch of 16 random noises is used to calculate the $\bm{x}_0$-loss every iteration. After each update, 8 random samples are generated. If any of them has a similarity with the target image no less than 0.5, the optimization process will stop.

\textbf{Calibration.}
SSCD similarity on the low-resolution CIFAR-10 yields many false positives~(non-replication with high similarity, early stopped) and false negatives~(replication with low similarity, not early stopped), although it works well on the other three high-resolution datasets. We perform a manual review on the false samples and correct their early stop timesteps and scores. All the other experimental results are unhandled.

\begin{figure}[t]
    \centering
    \includegraphics[width=\linewidth]{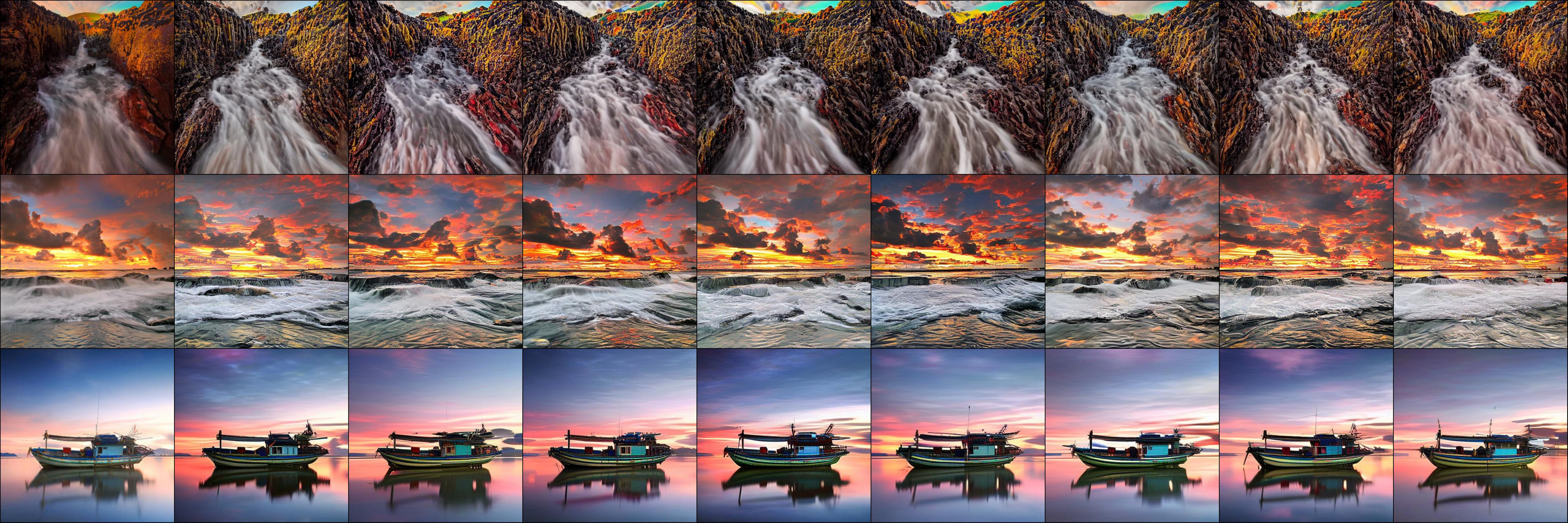}
    \caption{
      Not invertible examples in SD v1.4. The first column shows the corresponding training images.
    }
    \label{fig:not_invertible_examples}
\end{figure}

\textbf{Invertibility}. More than half of the images in CIFAR-10 are not invertible even if we set $\lambda=0$ all the time. This is probably because the DDPM~(34.21 M) trained on CIFAR-10 has limited capacity to memorize every training image. Images in CelebAHQ and FFHQ are all invertible. As a comparison, the LDMs trained on them have larger capacity~(314.12 M). There are also some images not invertible in Stable Diffusion. However, setting $\lambda=0$ could invert them almost perfectly. The invertibility of an image is specific to a set of hyperparameter settings. Although it is possible to invert everything in SD, we regard samples not invertible in our setting ``insignificant'', as compared to those that are easy to invert. \Cref{fig:not_invertible_examples} shows three examples that are not invertible in SD v1.4 from the \textit{normal} subset, as judged by SSCD. It shows that it is quite subjective to judge the similarity between images. Human may regard such cases as replication as the generated images mimic the style of training image or else. From a technical perspective, we leave this problem untouched and follow the decision of SSCD.

\begin{figure}[t]
    \centering
    \includegraphics[width=\linewidth]{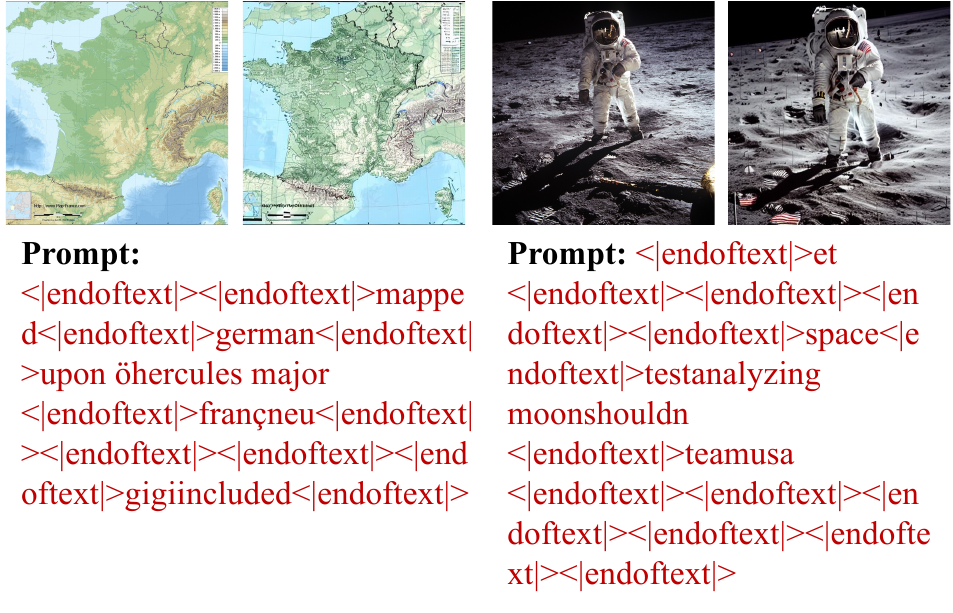}
    \caption{
      Additional examples of adversarial prompts found by GCG.
    }
    \label{fig:gcg_additional}
\end{figure}

\textbf{Adversarial prompts.}
\Cref{fig:gcg_additional} shows additional examples of adversarial prompts found by GCG for images from the suspicious set, which have lower InvMM scores.

\subsection{Collate InvMM with nearest neighbor test}

\begin{table*}
  \centering
  \scalebox{0.9}{
  \begin{tabular}{@{}lcccccc@{}}
    \toprule
    Setting & CelebAHQ-250 & CelebAHQ-2.5k & FFHQ-600 & FFHQ-6k & CelebAHQ & FFHQ \\
    \midrule
    Default & N/A(0) & N/A(0) & N/A(0) & N/A(0) & N/A(0) & N/A(0)\\
    Epoch $\times 2$ & 0.059(9) & N/A(0) & N/A(0) & N/A(0) & - & -\\
    Epoch $\times 3$ & 0.326(120) & N/A(0) & 0.000(1) & N/A(0) & - & -\\
    Epoch $\times 4$ & 0.567(181) & 0.000(3) & 0.000(4) & N/A(0) & - & -\\
    Duplicate $\times 5$ & - & 0.182(65) & - & 0.000(2) & - & -\\
    Duplicate $\times 10$ & - & 0.852(237) & - & 0.192(186) & - & -\\
    Duplicate $\times 15$ & - & 0.937(245) & - & 0.401(323) & - & -\\
    Duplicate $\times 20$ & - & 0.819(231) & - & 0.434(375) & - & -\\
    \bottomrule
  \end{tabular}
  }
  \caption{Performance of collating InvMM with nearest neighbor test. Each result refers to IoU(Number of replicated training images).}
  \label{tab:iou}
\end{table*}

We provide another quantitative metric to collate InvMM with the nearest-neighbor test: if a training image is replicated by its nearest neighbors in randomly generated samples, then its InvMM should be small among a list of training images, and vice versa. Let $S_{\textrm{nn}}$ be the set of images that expose replication in randomly generated samples, and $S_{\textrm{InvMM}}$ be the same-size set of images with the lowest InvMMs in a list of images. We define the consistence between InvMM and nearest-neighbor test as the Intersection over Union~(IoU) between $S_{\textrm{InvMM}}$ and $S_{\textrm{nn}}$:
\begin{equation}
    \mathrm{IoU} = \frac{|S_{\textrm{InvMM}}\cap S_{\textrm{nn}}|}{|S_{\textrm{InvMM}}\cup S_{\textrm{nn}}|},|S_{\textrm{nn}}|>0
\end{equation}

Under the setting in \cref{sec:detect}, InvMM achieves an IoU of 0.817 on CIFAR-10 and 1.0 on LAION. The results on CelebAHQ and FFHQ are summarized in \cref{tab:iou}. 10k random samples are used to obtain the results. A random sample of cosine similarity larger than 0.5 with any training sample is considered a replication. InvMM presents consistence with the results of nearest neighbor test, indicating its potential to expose risk of training image leakage, especially when a large number of training images are prone to leakage. The 10k samples are not an adequate sampling of the large latent space~(64$\times$64$\times$3), the performance can be refined with a larger scale of evaluation.

\subsection{Comparison to membership}
\label{sec:supp_membership}
\textbf{Experiment details.}
\Cref{fig:membership} is evaluated with LDM trained on FFHQ-6k. FFHQ-600 is used as the member set and another 600 images not contained in FFHQ-6k constitute the hold-out set. The statistic variant of SecMI is implemented following their official settings, with $t_{\mathrm{SEC}}=100$ and $k=10$.

\begin{figure}[t]
    \centering
    \includegraphics[width=\linewidth]{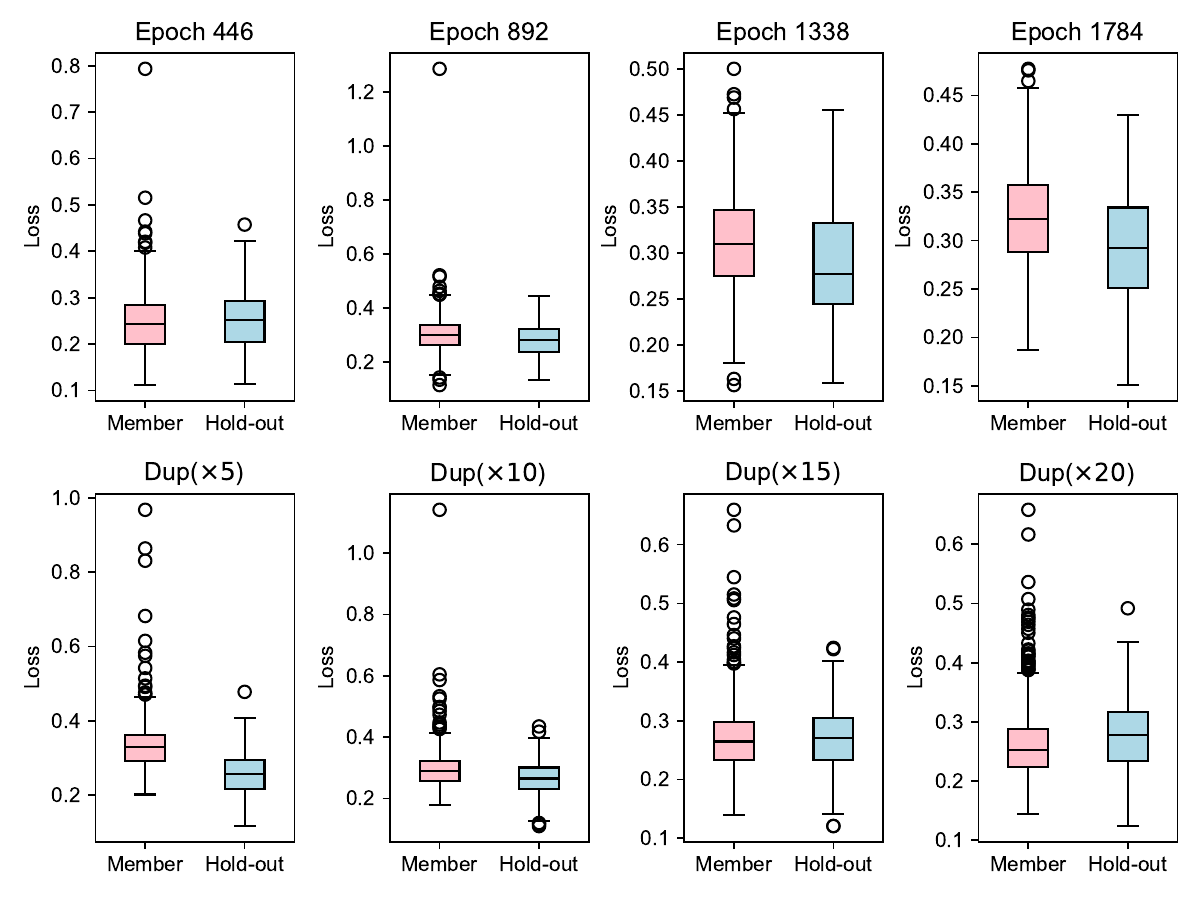}
    \caption{
    Comparison of replication loss for members and hold-out samples.
    }
    \label{fig:replication_loss}
\end{figure}

\begin{figure}[t]
    \centering
    \includegraphics[width=\linewidth]{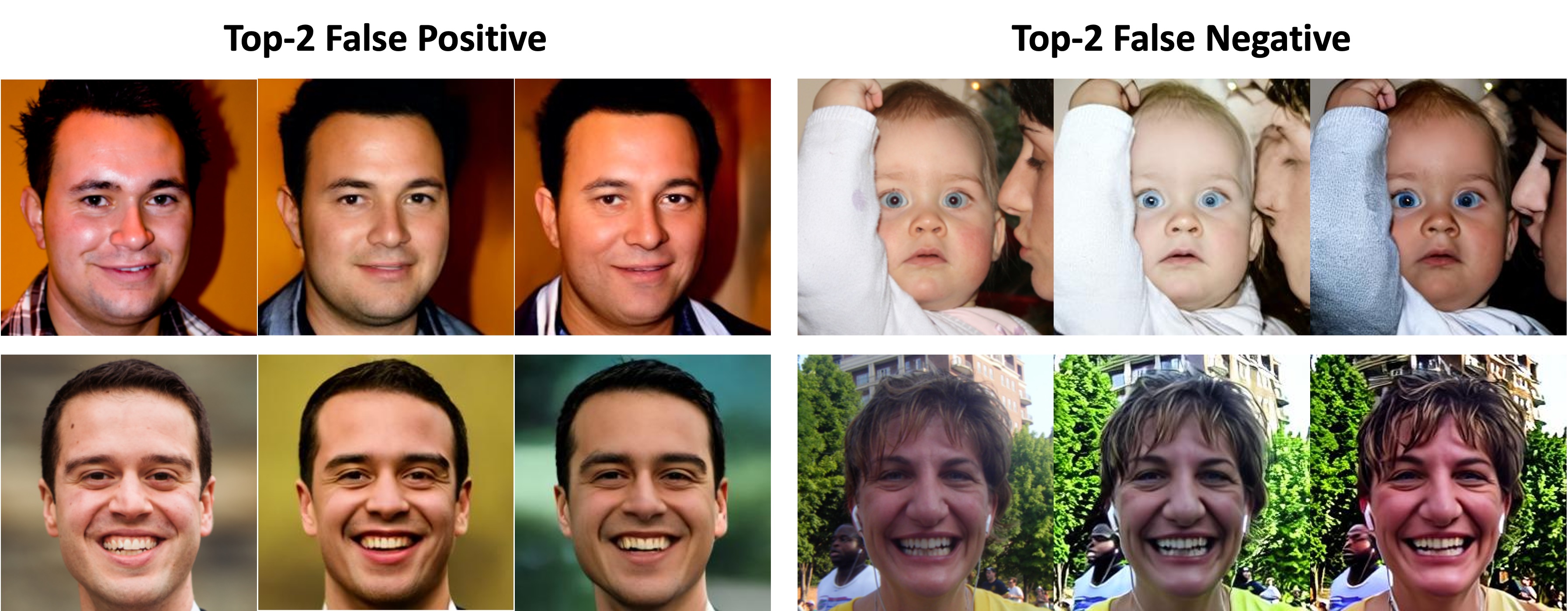}
    \caption{
    Top-2 false positives with lowest InvMM from the hold-out set and top-2 false negatives with highest InvMM from the member set. The results come from LDM trained on FFHQ-6k for 1784 epochs.
    }
    \label{fig:false_pos_neg}
\end{figure}

\textbf{Results.}
\Cref{fig:replication_loss} provides quantitative evidence that different samples could be replicated with different levels of training loss. The replication loss is calculated over the inverted latent noise distribution. It shows that although holdout samples have a larger training loss over the $\mathcal{N}(\mathbf{0},\mathbf{I})$, supporting effective membership inference, it is still possible to replicate some of them by reducing their training loss to a level higher than some of the members. 

We visualize the top-2 false positives and false negatives in \cref{fig:false_pos_neg}. The results show that hold-out samples are also replicated near identically.

\textbf{The result further highlights the difference between membership and memorization: training loss does not completely determine data replication.}

\section{Ablation study on prompt inversion}
\label{sec:text_to_image_inversion}
This section elaborates the influence of several factors for prompt inversion in SD, including the temperature $\tau$ in Gumbel-Softmax, the Classifier-Free Guidance~(CFG)~\cite{ho2022classifier} scale $\gamma$ and the advantage of predicting the image rather than noise for inversion~(\cref{eq:x0_prediction}).
For this goal, the noise distribution is temporarily fixed to the standard Gaussian. Experiment results will show that heavily memorized images can also be uncovered in this setting.

\begin{figure}[t]
    \centering
    \includegraphics[width=\linewidth]{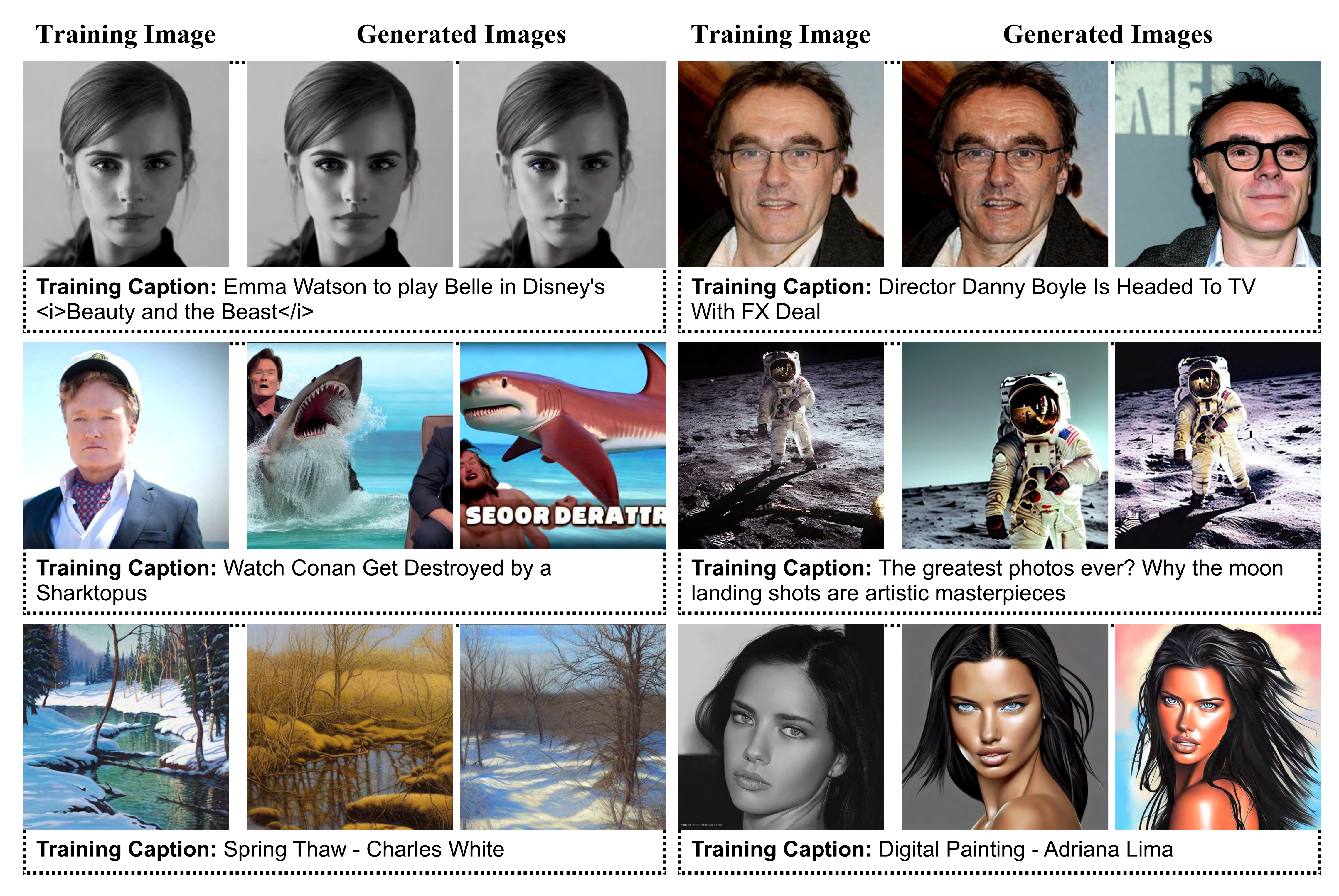}
    \caption{
      Examples in confirmed, suspicious and normal subsets from top to down. In each block, the right two columns show generated images using their training captions but different initial noises.
    }
    \label{fig:dataset_example}
\end{figure}

\subsection{Dataset}
We use the aforementioned three LAION subsets for evaluation. \Cref{fig:dataset_example} presents examples from the three subsets, together with images generated using their training captions.

\subsection{Experiment setup}
We utilize SD v1.4 for evaluation, which uses CLIP~\cite{radford2021learning} to encode input prompts and generate high-resolution $512\times512$ images. The text encoders by default accepts a maximum length of 77 tokens, in which the first and last tokens are padded tokens indicating the start and end of a prompt. The rest 75 tokens are all optimized in our experiments. During optimization, the parameters $\bm{\theta}$ of diffusion models are fixed. We optimize for 500 iterations with a constant learning rate of 0.2. $\bm{\phi}$ is initialized to $\mathbf{0}$ at the begining of optimization.

\subsection{Influence of temperature}
Hard prompt inversion to exactly reconstruct certain images is a challenging problem as it requires to search over a large and discrete space consisting of tens of thousands of tokens~(49408 in CLIP). We have found that the convergence of inversion relies on appropriate choice of the temperature $\tau$ in Gumbel-Softmax smoothing. With $\tau$ approaching 0, $\tilde{\bm{\omega}}$ drawn from the prompt distribution approaches one-hot and accurately matches a token, while it is difficult to optimize through gradient decent. Larger $\tau$ provides a smoother landscape of the target loss function and thus is easier to optimize. However, the smoothed $\tilde{\bm{\omega}}$ cannot directly correspond to some tokens. Discretizing them anyway~(\cref{eq:discretize}) might not preserve the same effectiveness as the smooth counterparts.

\begin{figure}[tbp]
    \centering
    \includegraphics[width=0.7\linewidth]{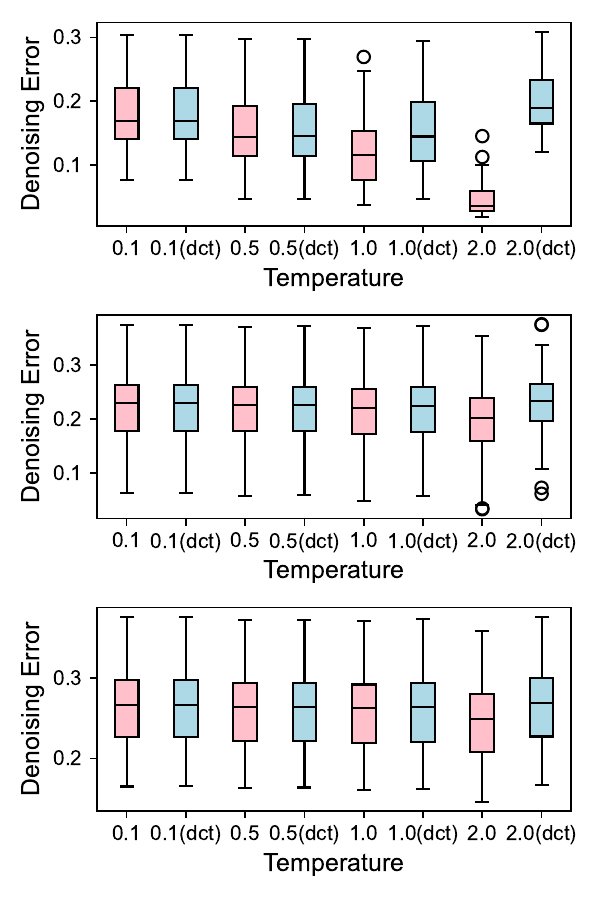}
    \caption{The distribution of denoising error of the confirmed, suspicious and normal subsets. "dct" means plus discretization.}
    \label{fig:denoising_error}
\end{figure}

\textbf{Denoising error.}
We first analyze the denoising error of the inverted prompt distribution when optimized with different temperatures. We consider 4 settings for the temperature to be either 0.1, 0.5, 1 or 2, as well as whether to discretize the smoothed tokens $\tilde{\bm{\omega}}$ to $\bm{\omega}$.
After optimization, random prompts and noises can be drawn from $q_{\bm{\phi}}(\bm{\omega}|\bm{x}_0)$ and $\mathcal{N}(\mathbf{0},\mathbf{I})$. For each image, we randomly sample 10 prompts and 10 noises for each sampled prompt, resulting in 100 prompt-noise pairs. The denoising error is estimated using 50 timesteps uniformly sampled within the range $[1,1000]$ and averaged over the 100 prompt-noise pairs.

The results can be seen in ~\cref{fig:denoising_error}. For any type of image from different groups, higher temperatures lead to lower denoising errors on average, indicating a more adequate optimization. However, meanwhile, plus token discretization worsens the effectiveness.

\begin{figure*}[tbp]
    \centering
    \begin{subfigure}[t]{0.31\linewidth}
        \centering
        \includegraphics[width=\linewidth]{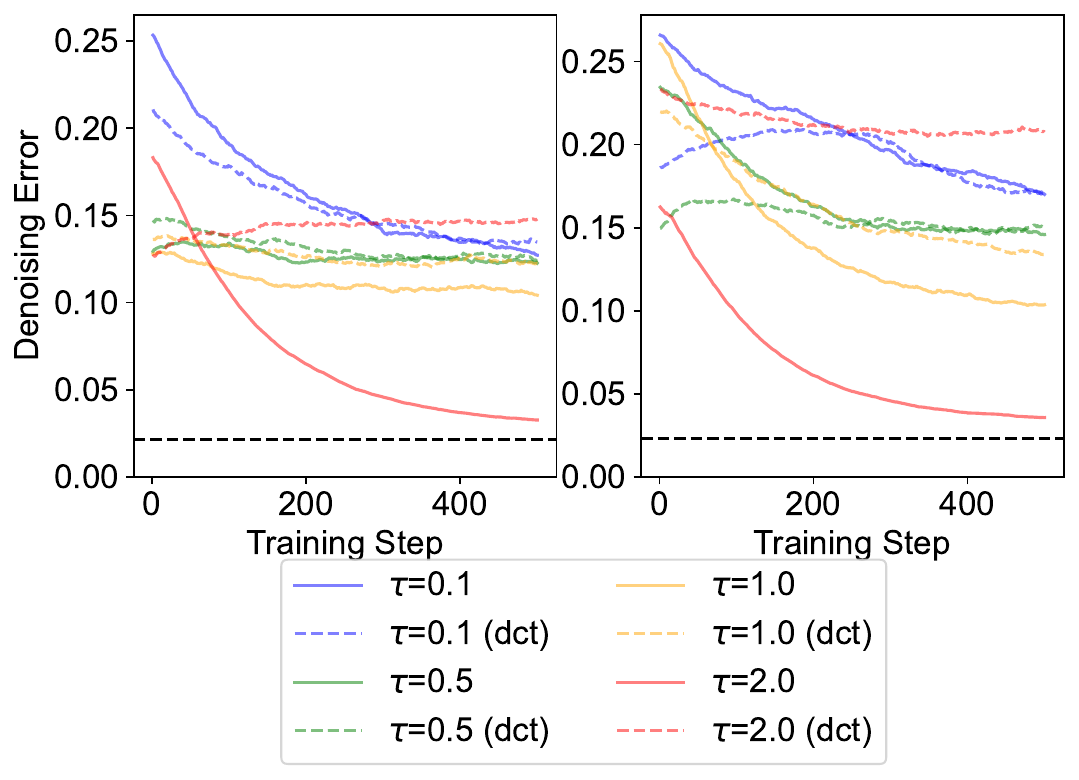}
        \caption{Confirmed Examples}
        \label{fig:convergence_example_confirmed}
    \end{subfigure}
    \begin{subfigure}[t]{0.31\linewidth}
        \centering
        \includegraphics[width=\linewidth]{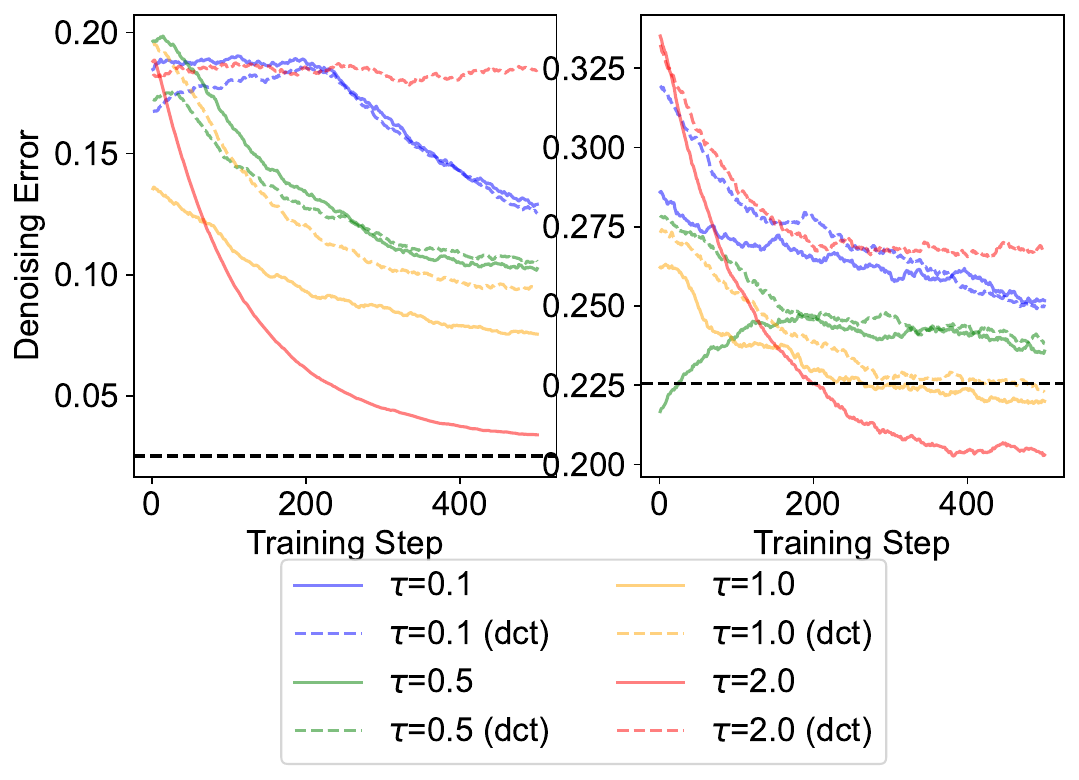}
        \caption{Suspicious Examples}
        \label{fig:convergence_example_suspects}
    \end{subfigure}
    \begin{subfigure}[t]{0.31\linewidth}
        \centering
        \includegraphics[width=\linewidth]{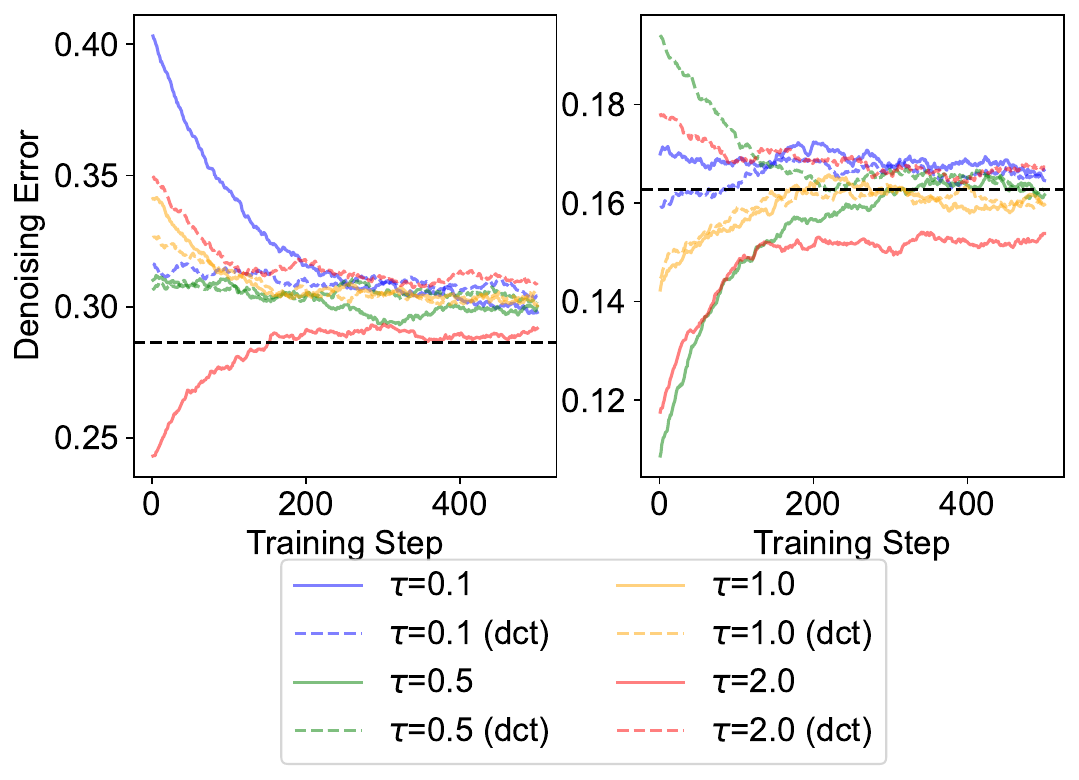}
        \caption{Normal Examples}
        \label{fig:convergence_example_normal}
    \end{subfigure}
    \caption{Training denoising error of examples from each group under different temperatures, smoothed via exponential moving average with a momentum of 0.99. "dct" means plus discretization. The black dashed line is the baseline error calculated using their training captions over 1600 randomly sampled Gaussian noises.}
    \label{fig:convergence_example}
\end{figure*}

\begin{figure}[t]
    \centering
    \includegraphics[width=\linewidth]{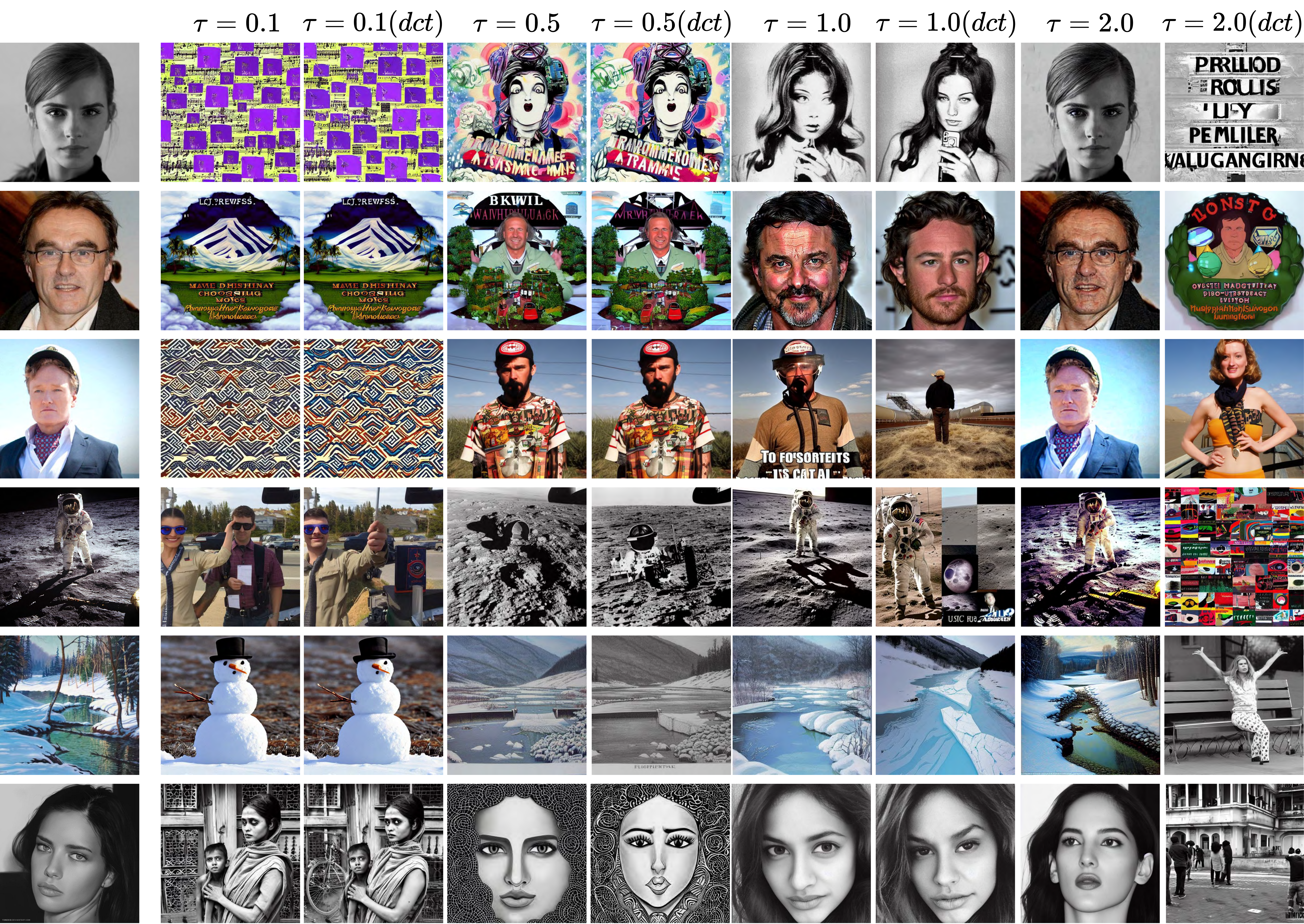}
    \caption{
      Generation results of different temperatures. The first column shows the corresponding training images.
    }
    \label{fig:tau_generation_results}
\end{figure}

\textbf{Convergence.}
~\Cref{fig:convergence_example} shows the denoising errors at each optimization step of the 6 example images in ~\cref{fig:dataset_example}. For the assessment of convergence, we draw a baseline denoising error calculated using the training caption of each image. As can be observed, large temperatures induce better convergence and the difference between $\tilde{\bm{\omega}}$ and $\bm{\omega}$ becomes prominent. \Cref{fig:tau_generation_results} illustrates the generation results using prompts sampled from the learned distribution, with a CFG scale of 7. As can be seen, with $\tau=2$, the inverted prompts are able to replicate the training images for the 4 examples from the confirmed and suspicious groups, the two from the suspicious group are newly found through our analysis. However, $\tau=2$ plus discretization produces completely irrelevant images. Lower temperatures 0.5 and 1.0 present more consistent generation between smooth and discrete prompts, while they only produce similar images to the training ones, showing analogous content, color, etc. The smallest $\tau=0.1$ fails to capture the main content of the training images but remains the best consistency for discretization.

\begin{figure}[t]
    \centering
    \includegraphics[width=0.6\linewidth]{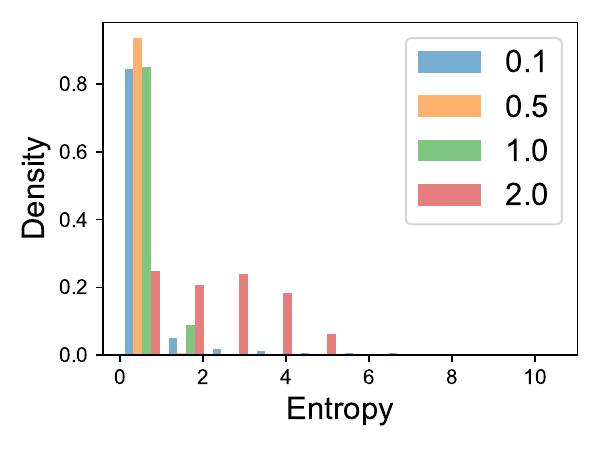}
    \caption{
    Entropy density distribution of the prompt categorical distribution.
    }
    \label{fig:distribution}
\end{figure}

\textbf{Prompt distribution.}
\Cref{fig:distribution} depicts the density distribution of the entropy $-\sum_{j=1}^{|\mathcal{V}|} \bm{\pi}_{i,j}\log  \bm{\pi}_{i,j}$ of the learned prompt categorical distributions. When $\tau=2$, most tokens follow a high entropy distribution, which means that they are well smoothed and take an interpolation of hard tokens. In contrast, smaller temperatures produce more sharp distributions, while less effective as large temperature for inversion.

\textbf{Conclusion.}
For the goal of effective analysis, we adopt a compromise setting with the temperature $\tau$ of 2.0 and without discretization, to reach adequate optimization. Although this violates the goal of inverting realistic prompts, it is reasonable and enough for developers to analyze the vulnerability of their models. Note that it still offers a certain level of restriction to the learned soft prompts, as the Gumbel-Softmax approximation together with a linear combination of pretrained token embeddings bound the smoothed tokens $\tilde{\bm{\omega}}$ in the convex hull of the pretrained tokens. 

\subsection{CFG Scale}
\begin{figure}[t]
    \centering
    \includegraphics[width=\linewidth]{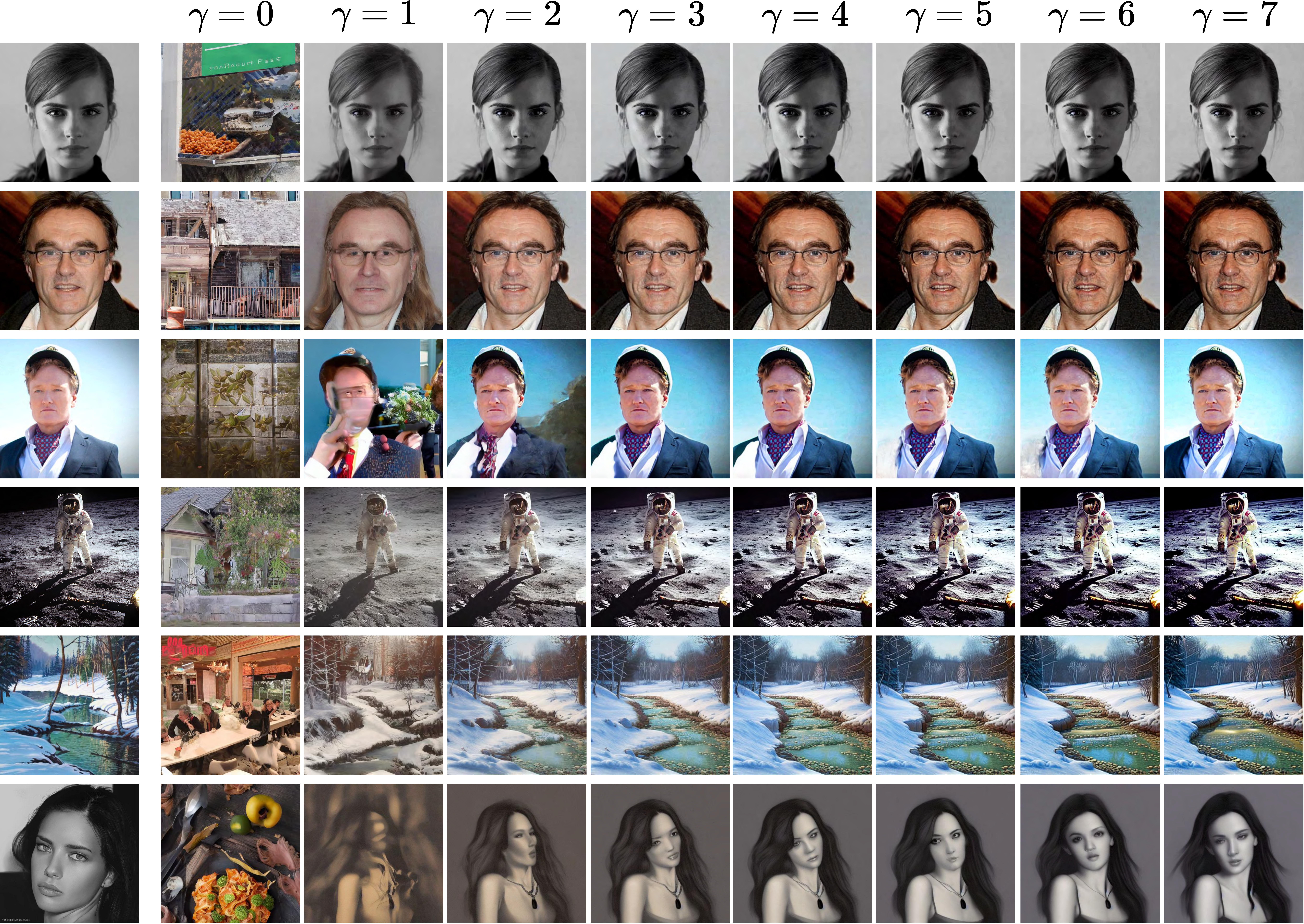}
    \caption{
      Generation results of different classifier-free guidance scales. The first column shows the corresponding training images.
    }
    \label{fig:cfg_scale}
\end{figure}

In additional, we sweep the CFG scale $\gamma$ from 0 to 7 with interval 1 to study its influence. $\gamma=0$ indicates unconditional generation and $\gamma=1$ indicates conditional generation without penalizing unconditional prediction. The generation results of the examples in \cref{fig:dataset_example} are shown in \cref{fig:cfg_scale}.

It can be observed that the generated images with $\gamma=0$ are quite random because they only depend on the random initial noises.
When $\gamma\geq 1$, for heavily memorized images in the confirmed and suspicious group, the generation results progressively converge to the training images. At times the generated images with small CFG scale only resemble the training images but are not eidetic, e.g., the 2rd to 4th rows. Nonetheless, we also discovered perfect replication for these examples with a small $\gamma=1$ generated using other different initial noises. This indicates that the extent to which different training images are memorized varies, and, moreover, a relatively low-level training time memorization~($\gamma=1$) can be amplified by sampling-time options such as larger $\gamma$.
Given that we optimize the prompts w.r.t. the conditional model~($\gamma=1$), it demonstrates that training data leakage roots in the conditional model.

In addition, a gradual sharpening can be observed in the generated images as the guidance scale increases. As we optimize w.r.t. the conditional model, i.e., $\gamma=1$, it is of enough denosing accuracy to generate an training image with relatively lower scales. Enlarging the conditional scale, however, results in excessive alignment with the input prompt. In contrast, for the images in the normal group~(see the last two rows of \cref{fig:cfg_scale}), as the inverted prompt distribution cannot fully capture its complete content, generation with $\gamma=1$ is somewhat fuzzy. It thus benefits from an increase of $\gamma$ for higher quality.

\textbf{Conclusion.}
Training data memorization can be amplified by CFG scale. As we consider the worst-case memorization in this paper, we count in the replication caused by any CFG scale from 1 to 7.

\subsection{Optimization Objective}
\begin{figure}[t]
    \centering
    \includegraphics[width=\linewidth]{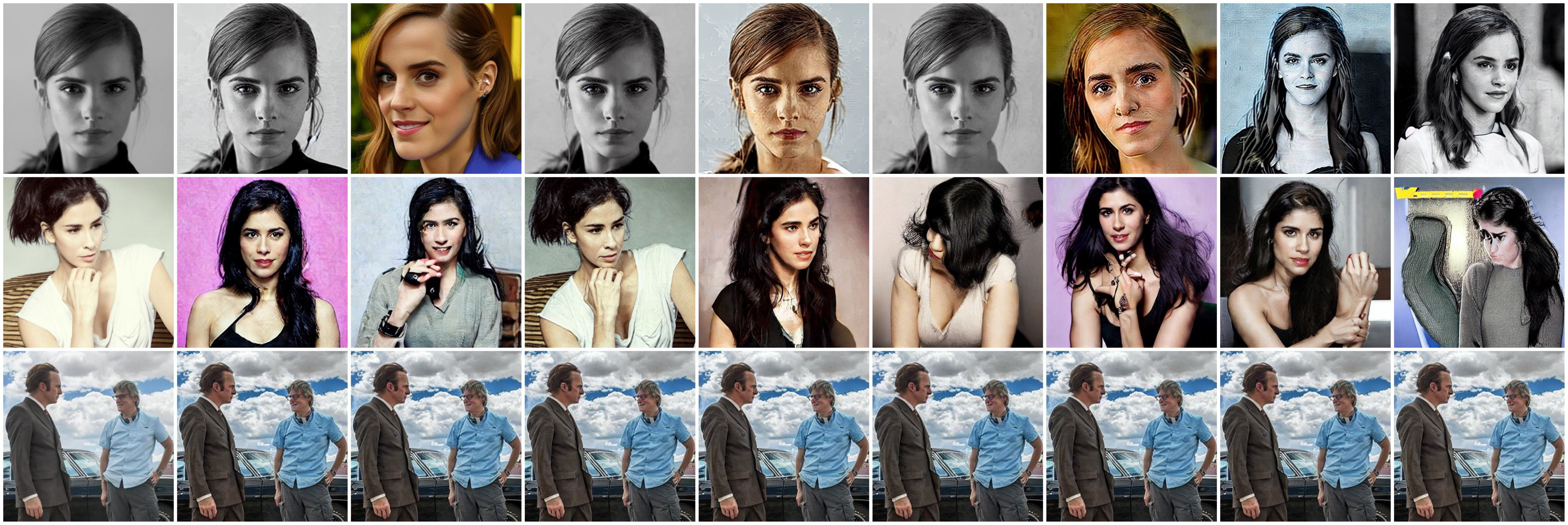}
    \caption{
      Inversion results of $\bm{\epsilon}_0$-prediction. The first column shows the corresponding training images.
    }
    \label{fig:epsilon-inversion}
\end{figure}

As we adopt the modified $\bm{x}_0$-prediction objective different from the original $\bm{\epsilon}_0$-prediction objective of the diffusion models used in our experiments, we verify the effectiveness of $\bm{x}_0$-prediction over $\bm{\epsilon}_0$-prediction for inversion. We evaluate using the images in the \textit{confirmed} set to determine if the $\bm{\epsilon}_0$-prediction can successfully replicate them. \Cref{fig:epsilon-inversion} shows the inversion results of $\bm{\epsilon}_0$-prediction. Inverison with $\bm{\epsilon}_0$-prediction is much unstable compared to $\bm{x}_0$-prediction, which demonstrates the importance of reweighting denoising error at different timesteps. More specifically, the later timesteps at training time~(ealier at sampling time) tend to shape the large scale image features~\cite{ho2020denoising}, e.g., shape, object. Therefore, it would be beneficial to upweight the later timesteps by $\bm{x}_0$-prediction to more accurately guide diffusion models to generate the corresponding training images.

\textbf{Conclusion.}
Although not aligning with the original training objective, $\bm{x}_0$-prediction is more stable for inversion.

\subsection{Adaptive algorithm}
\begin{figure}[t]
    \centering
    \includegraphics[width=0.8\linewidth]{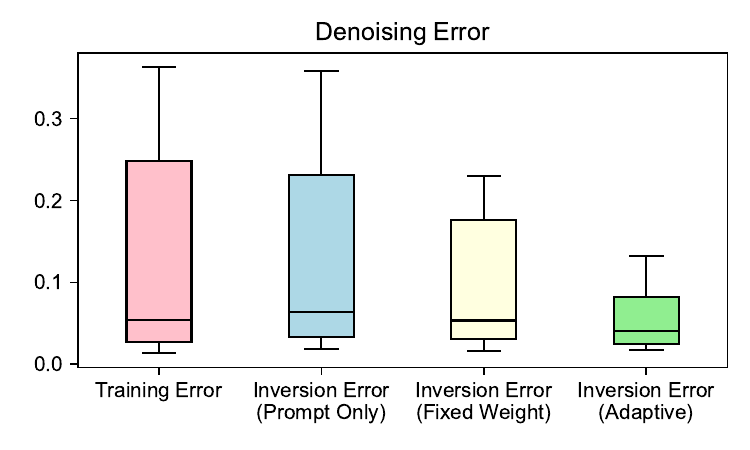}
    \caption{
    Comparison of inversion denoising error under different settings on SD v1.4.
    }
    \label{fig:sd_ablation_de_cmp}
\end{figure}
A comparison of the (1)~training error~(using training captions), (2)~inversion error with only prompt distribution learned, (3)~inversion error with both prompt and noise distributions learned, fixing $\lambda=1$ and (4)~inversion error with both prompt and noise distributions learned, dynamically adjusting $\lambda$ by \cref{alg:measure}, is shown in ~\cref{fig:sd_ablation_de_cmp}.
Compared to the training error, (2) only reduces that of heavily memorized images for which the input prompts plays a crucial role. (3) further reduces the denoising error but cannot work for all samples. \cref{alg:measure} can successfully reduce the denoising error of any training samples by adaptively adjusting the weight of normality regularization.

\section{More generation results}
\label{sec:supp_generation_results}
\Cref{fig:sdv1_confirmed_generation_results,fig:sdv1_suspects_generation_results,fig:sdv1_normal_generation_results,fig:sdv2_confirmed_generation_results,fig:sdv2_suspects_generation_results,fig:sdv2_normal_generation_results,fig:sdv3_confirmed_generation_results,fig:sdv3_suspects_generation_results,fig:celebahq_generation_results,fig:ffhq_generation_results} show more generated images using inverted noise vectors~(and prompts).


\begin{figure*}[t]
    \centering
    \includegraphics[width=\linewidth]{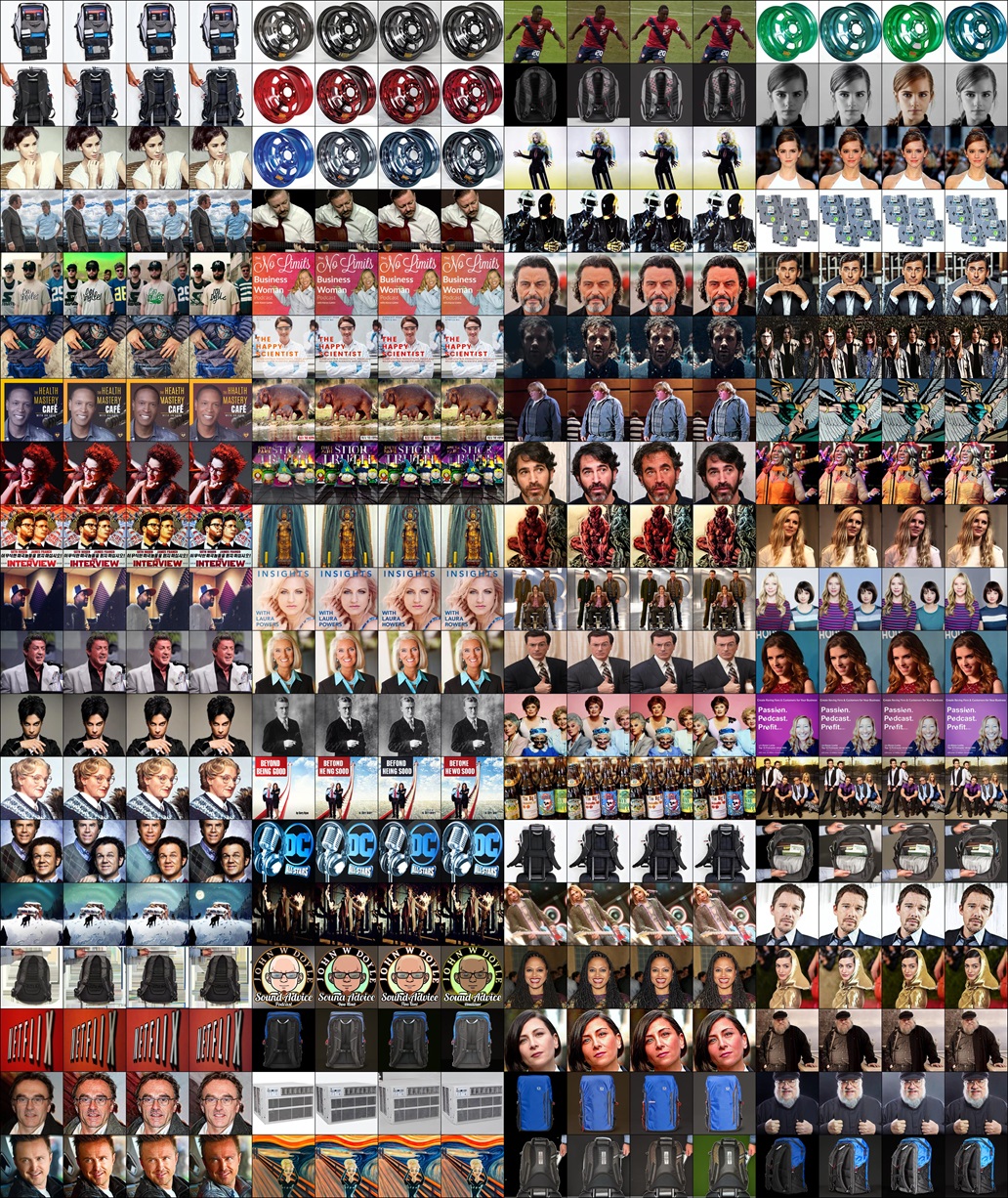}
    \caption{
      Random samples of SD v1.4 inversion on the \textit{confirmed} subset. The first column shows the corresponding training images.
    }
    \label{fig:sdv1_confirmed_generation_results}
\end{figure*}

\begin{figure*}[t]
    \centering
    \includegraphics[width=\linewidth]{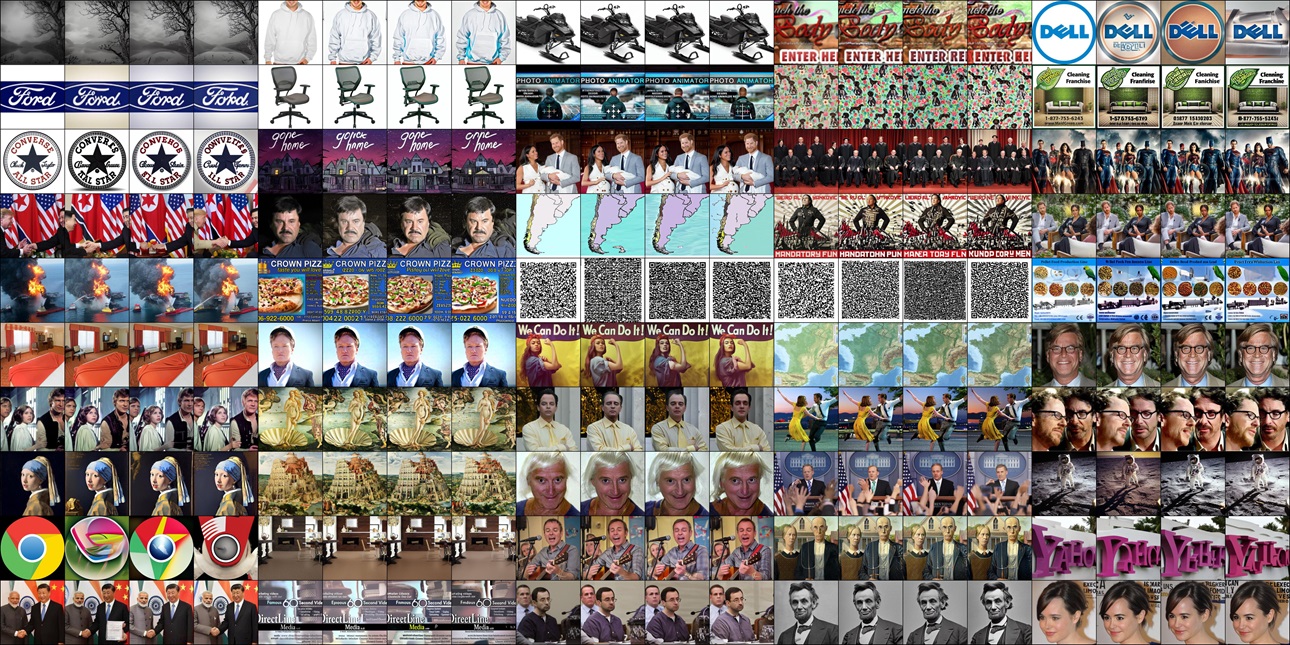}
    \caption{
      Random samples of SD v1.4 inversion on the \textit{suspicous} subset. The first column shows the corresponding training images.
    }
    \label{fig:sdv1_suspects_generation_results}
\end{figure*}

\begin{figure*}[t]
    \centering
    \includegraphics[width=\linewidth]{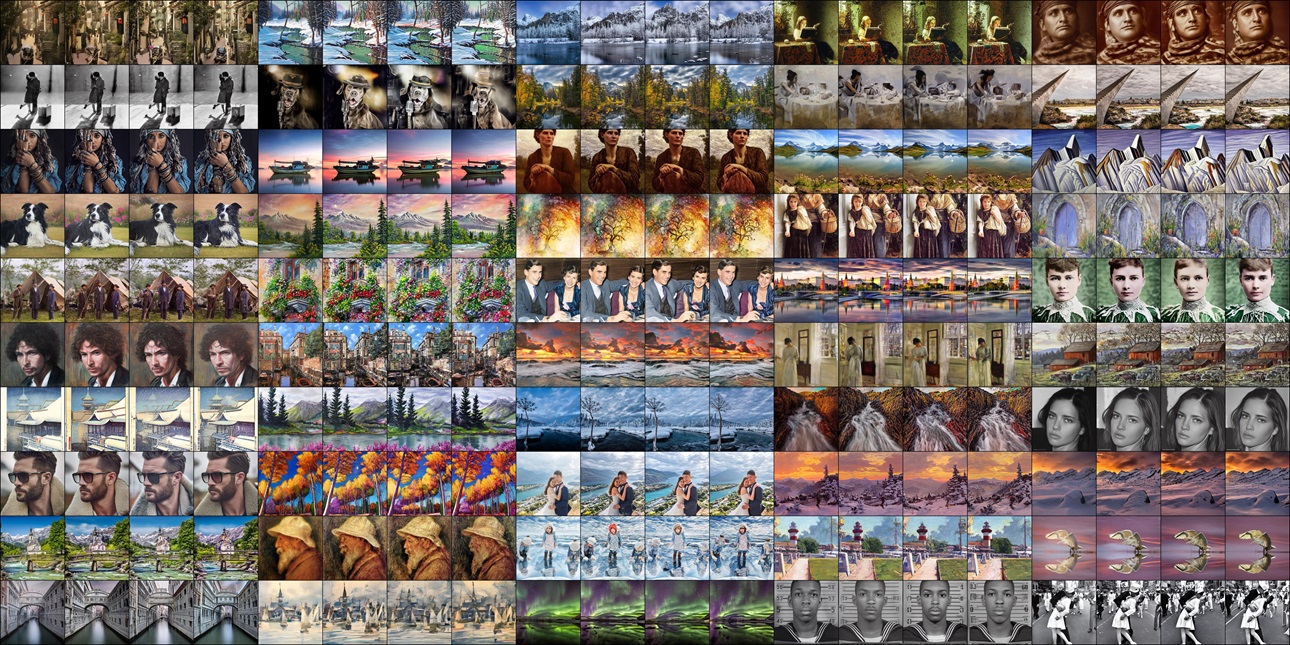}
    \caption{
      Random samples of SD v1.4 inversion on the \textit{normal} subset. The first column shows the corresponding training images.
    }
    \label{fig:sdv1_normal_generation_results}
\end{figure*}


\begin{figure*}[t]
    \centering
    \includegraphics[width=\linewidth]{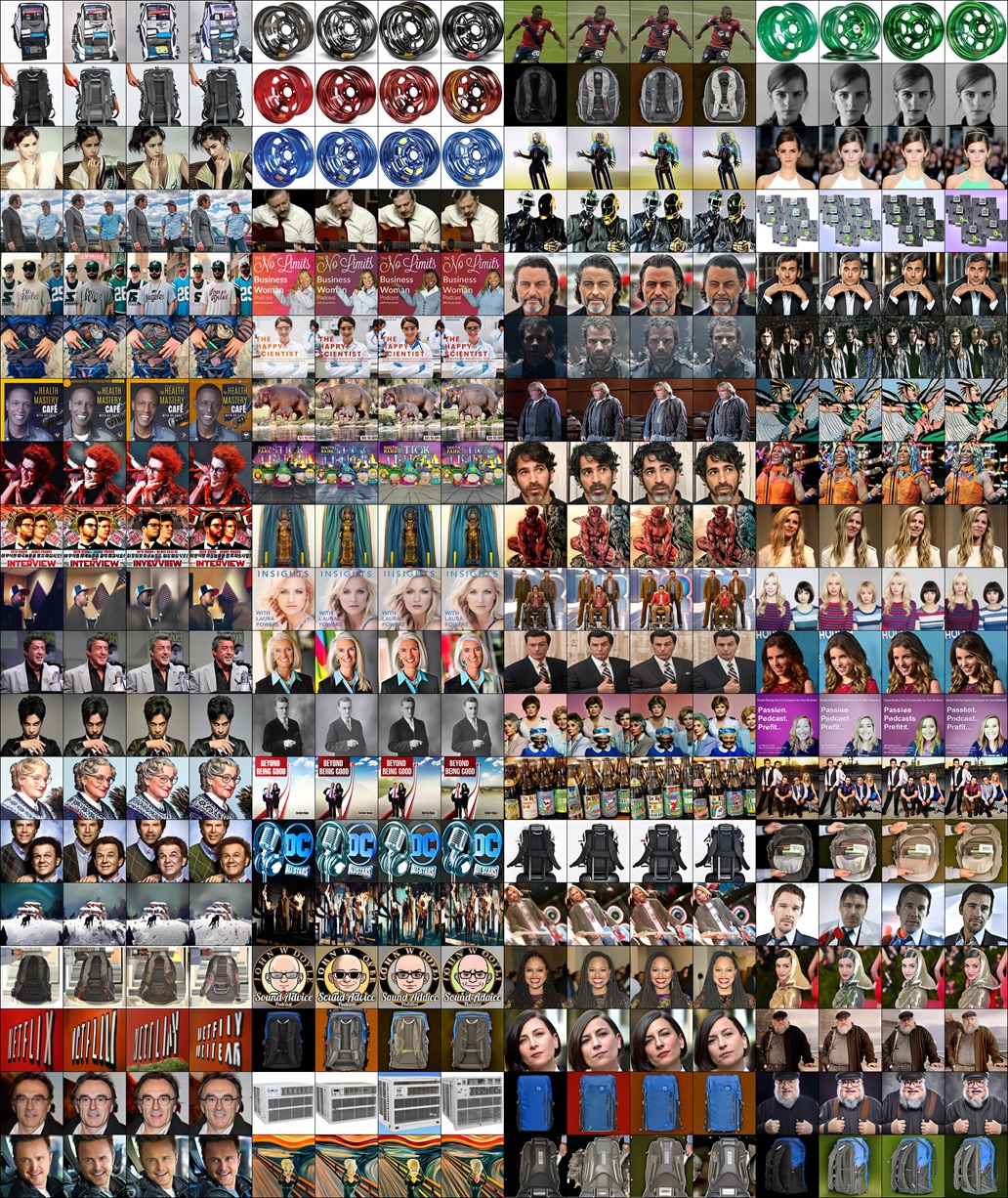}
    \caption{
      Random samples of SD v2.1 inversion on the \textit{confirmed} subset. The first column shows the corresponding training images.
    }
    \label{fig:sdv2_confirmed_generation_results}
\end{figure*}

\begin{figure*}[t]
    \centering
    \includegraphics[width=\linewidth]{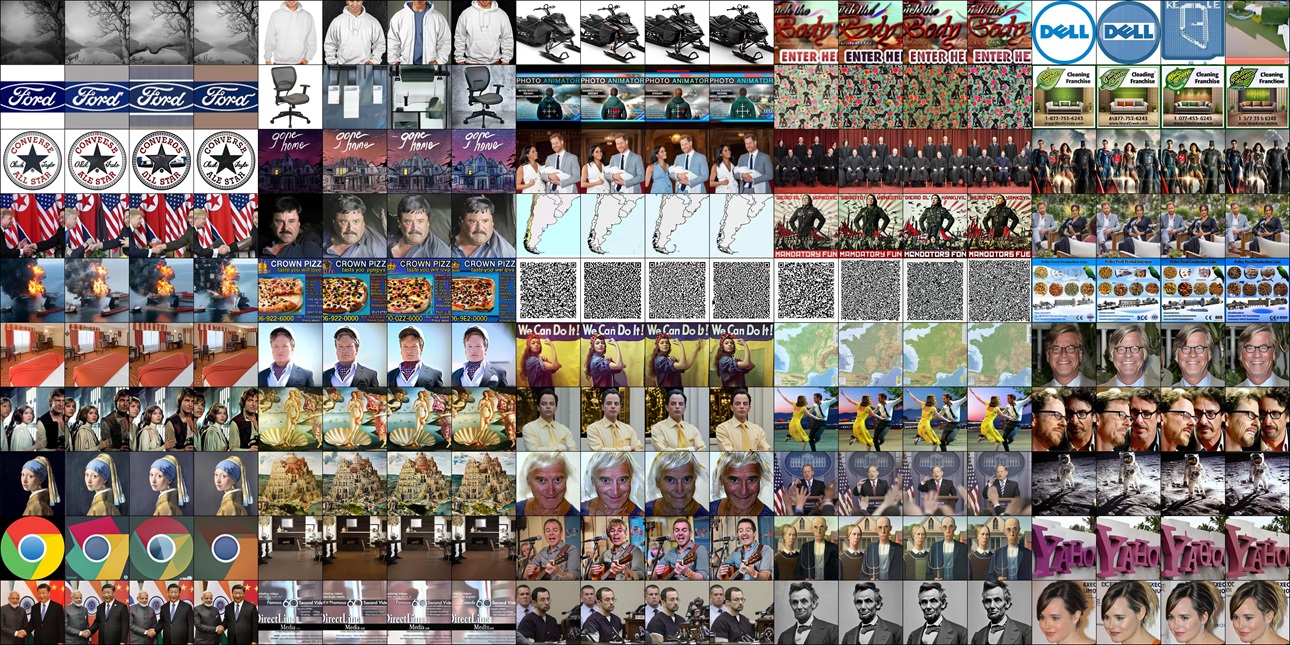}
    \caption{
      Random samples of SD v2.1 inversion on the \textit{suspicous} subset. The first column shows the corresponding training images.
    }
    \label{fig:sdv2_suspects_generation_results}
\end{figure*}

\begin{figure*}[t]
    \centering
    \includegraphics[width=\linewidth]{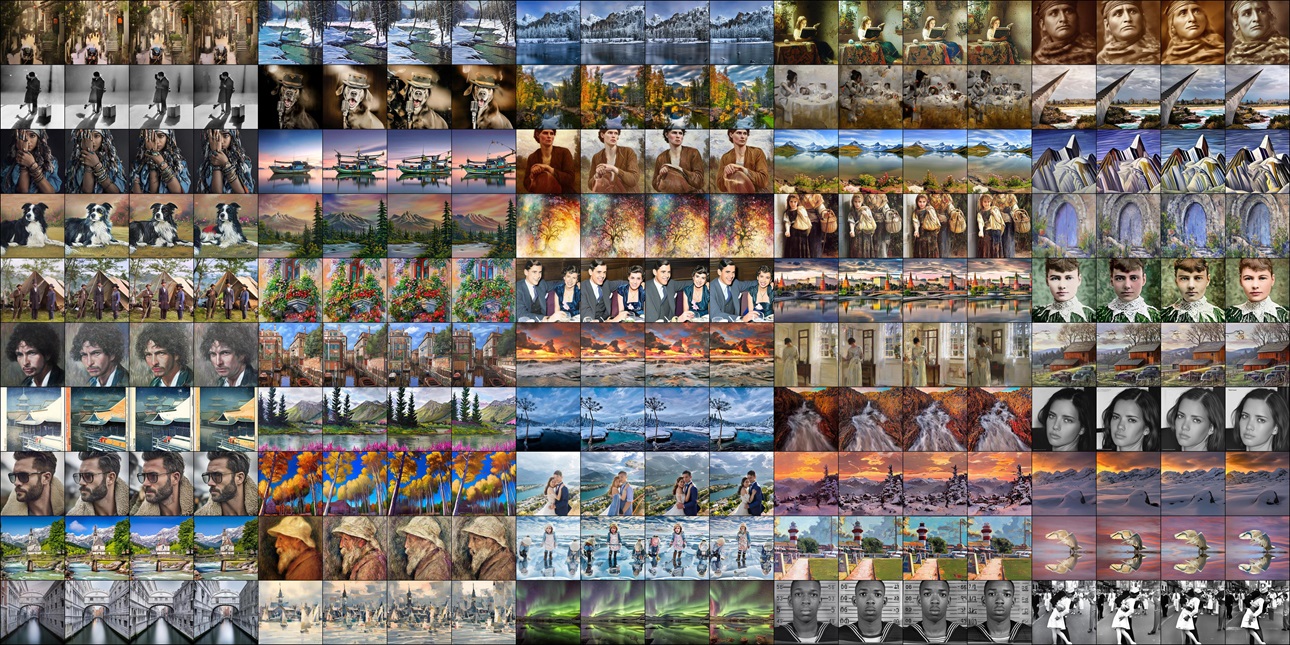}
    \caption{
      Random samples of SD v2.1 inversion on the \textit{normal} subset. The first column shows the corresponding training images.
    }
    \label{fig:sdv2_normal_generation_results}
\end{figure*}


\begin{figure*}[t]
    \centering
    \includegraphics[width=\linewidth]{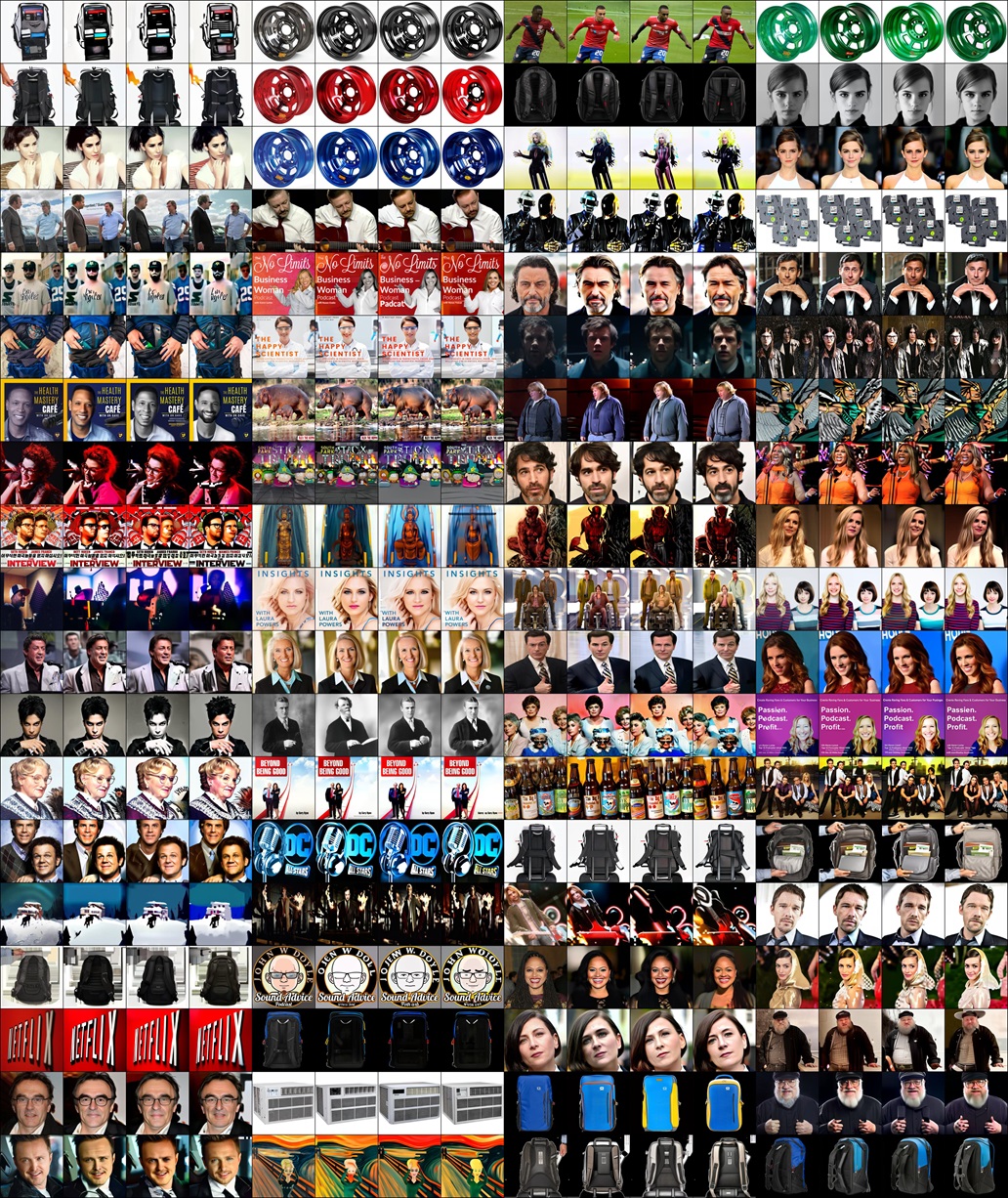}
    \caption{
      Random samples of SD v3.5 inversion on the \textit{confirmed} subset. The first column shows the corresponding training images.
    }
    \label{fig:sdv3_confirmed_generation_results}
\end{figure*}

\begin{figure*}[t]
    \centering
    \includegraphics[width=\linewidth]{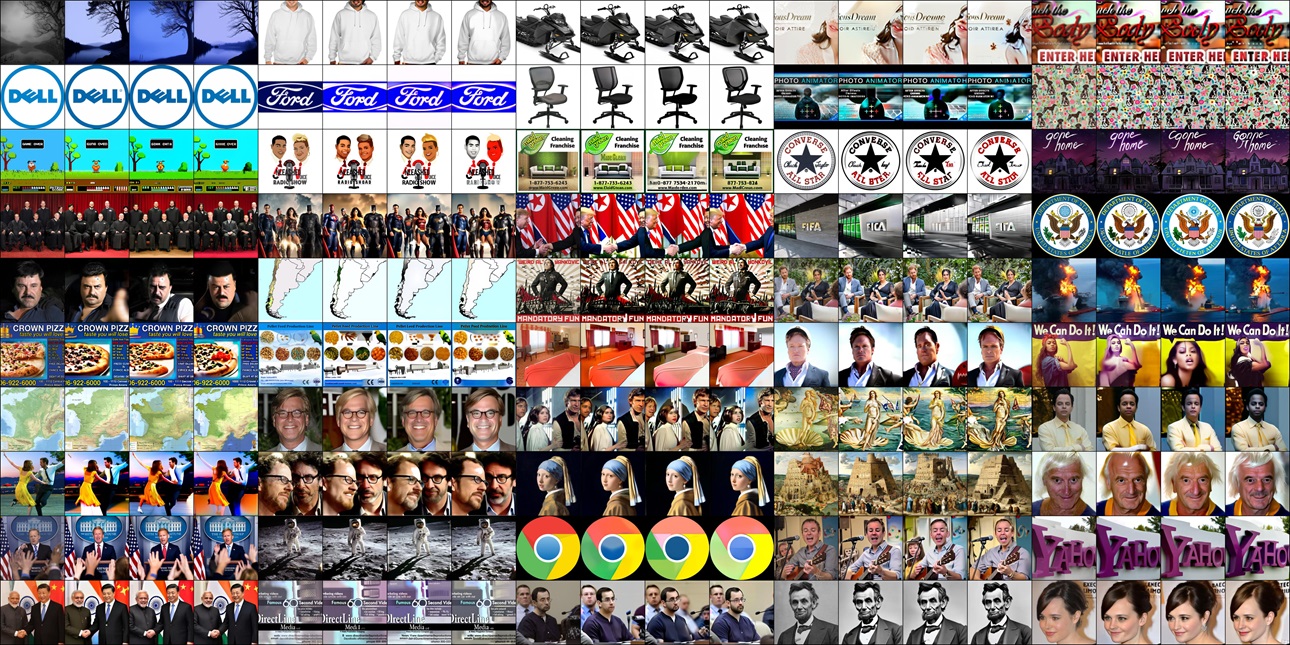}
    \caption{
      Random samples of SD v3.5 inversion on the \textit{suspicous} subset. The first column shows the corresponding training images.
    }
    \label{fig:sdv3_suspects_generation_results}
\end{figure*}

\begin{figure*}[t]
    \centering
    \includegraphics[width=\linewidth]{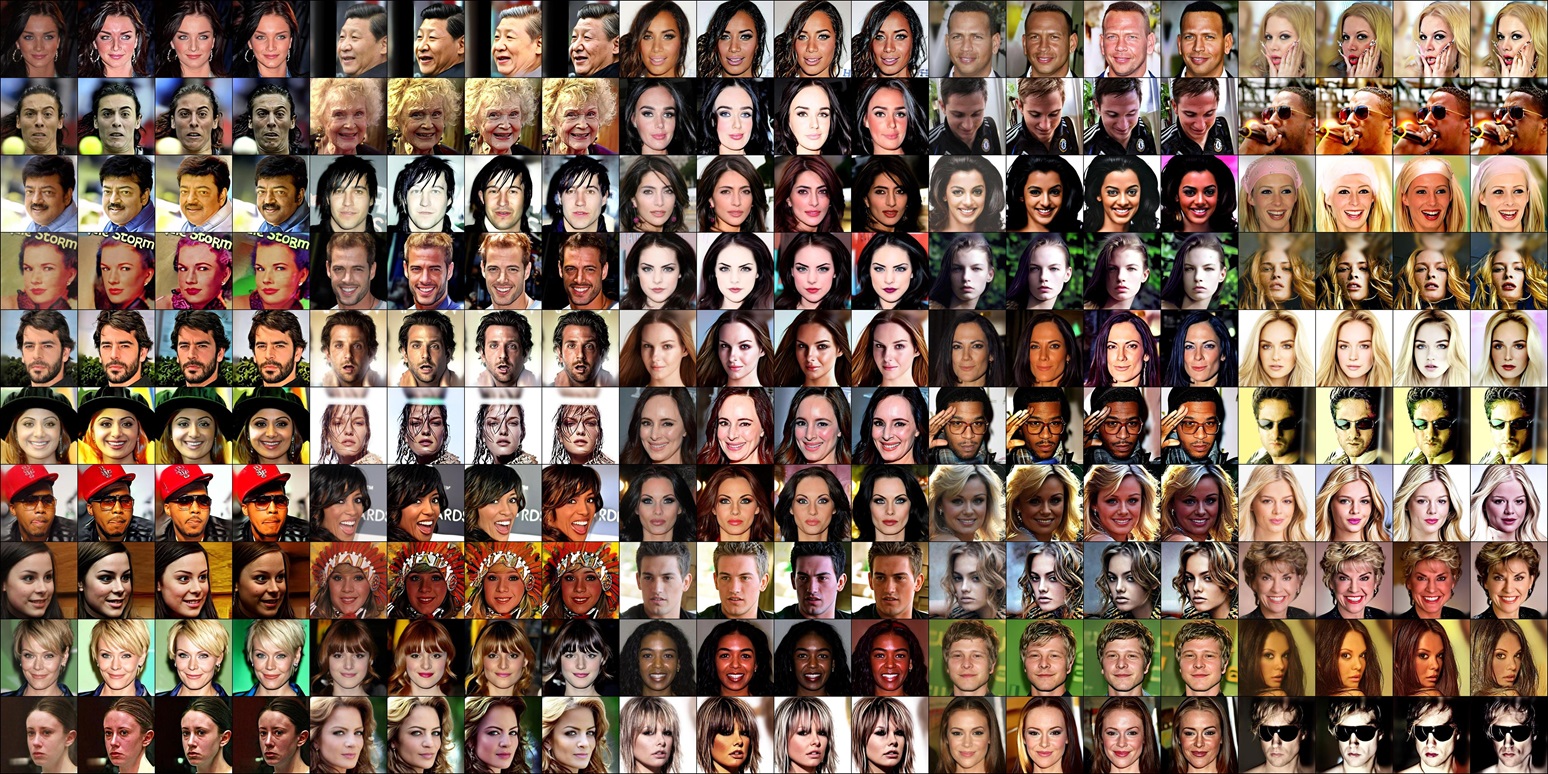}
    \caption{
      Random samples of LDM inversion on CelebAHQ. The first column shows the corresponding training images.
    }
    \label{fig:celebahq_generation_results}
\end{figure*}

\begin{figure*}[t]
    \centering
    \includegraphics[width=\linewidth]{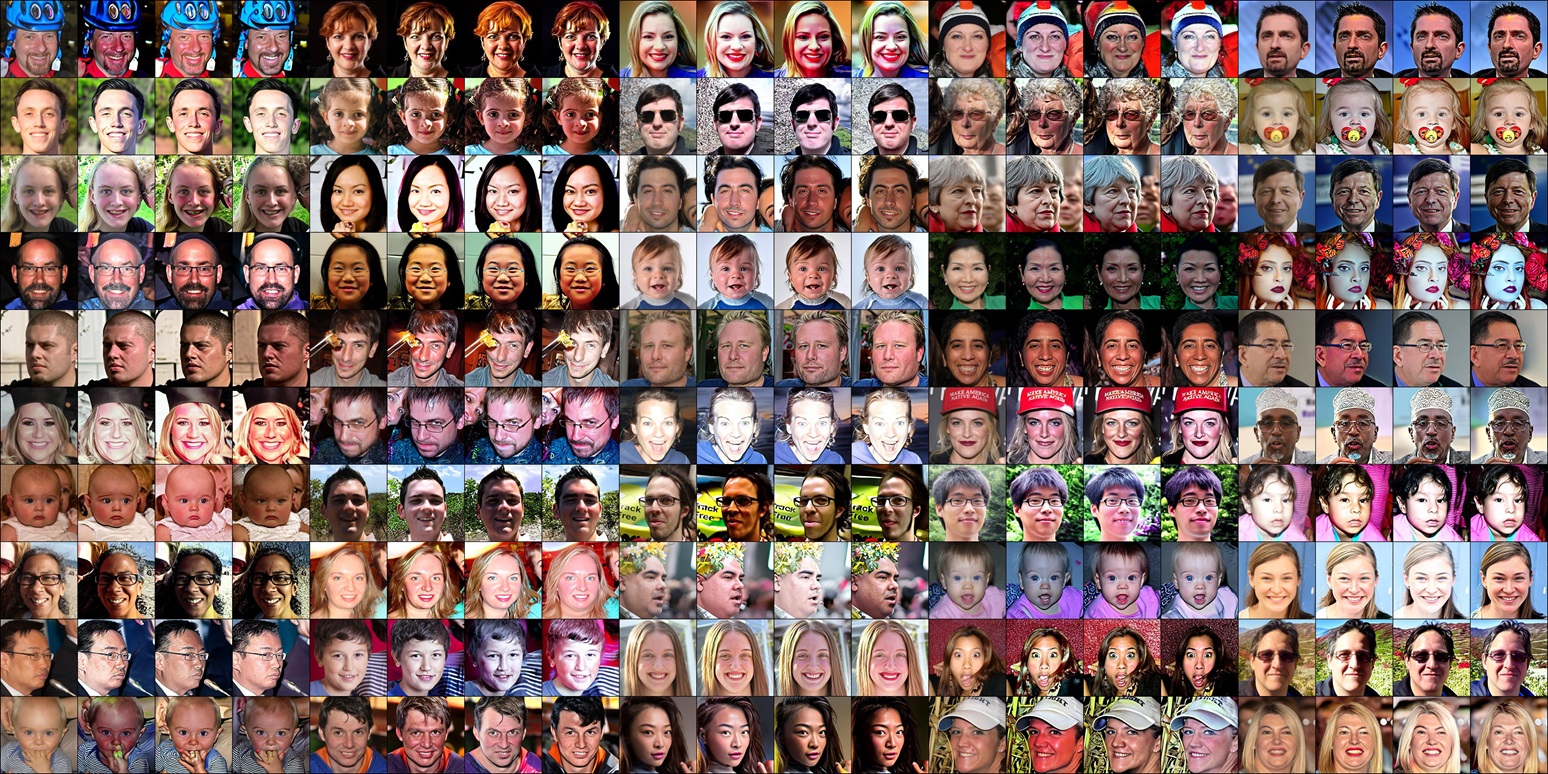}
    \caption{
      Random samples of LDM inversion on FFHQ. The first column shows the corresponding training images.
    }
    \label{fig:ffhq_generation_results}
\end{figure*}

\end{document}